\documentclass[prd,twocolumn,showpacs,eqsecnum,floatfix,preprintnumbers]{revtex4}
\usepackage{graphicx}
\usepackage{dcolumn}
\usepackage{bm}

\usepackage{amsmath}
\usepackage{graphicx}
\usepackage{bbm}
\usepackage{color}
\usepackage{bm}
\usepackage{latexsym}
\usepackage{amssymb}
\usepackage{booktabs}
\usepackage[ansinew]{inputenc}
\usepackage{rotating}
\usepackage{euscript}
\usepackage{gensymb} 
\usepackage{subfigure}

\newcommand{\be}{\begin{equation}}

\begin{document}
\preprint{ADP-11-13/T735}

\title{Gluonic profile of the static baryon at finite temperature.}

\author{Ahmed S. Bakry*}
\affiliation{Special Research Center for the Subatomic Structure of Matter, Department of Physics, University of Adelaide, South Australia 5005, Australia.}\email[]{abakry@physics.adelaide.edu.au}

\author{Derek B. Leinweber}%
\affiliation{Special Research Center for the Subatomic Structure of Matter, Department of Physics, University of Adelaide, South Australia 5005, Australia.}

\author{Anthony G. Williams}
\affiliation{Special Research Center for the Subatomic Structure of Matter, Department of Physics, University of Adelaide, South Australia 5005, Australia.}

\date{16 June 2011}
\begin{abstract}
   The gluon flux distribution of a static three quark system  has been revealed at finite temperature in the pure SU(3) Yang-Mills theory. An action density operator is correlated with three Polyakov loops representing the baryonic state at a temperatures near the end of the QCD plateau, $T/T_{c} \approx 0.8$, and another just before the deconfinement point, $T/T_{c} \approx 0.9$. The flux distributions at short distance separations between the quarks display an action-density profile consistent with a rounded filled $\Delta$ shape iso-surface. However the $\Delta$ shape action iso-surface distributions are found to persist even at large inter-quark separations. The action density distribution in the quark plane exhibits a nonuniform pattern for all quark separations considered. This result contrasts the well-known Y-shaped uniform action density gluonic-flux profile obtained using the Wilson loop as a quark source operator at zero temperature. We systematically measure and compare the main aspects of the profile of the flux distribution at the two considered temperature scales for three sets of isosceles triangle quark configurations. The radii, amplitudes and rate of change of the width of the flux distribution are found to reverse their behavior as the temperature increases from the end of the QCD plateau towards the deconfinement point. Remarkably, we find the mean square width of the flux distribution shrinks and localizes for quark separations larger than $1.0$ fm at $T/T_{c} \approx 0.8$ which results in an identifiable Y-shaped radius profile. Near the deconfinement point, the action-density delocalizes and the width broadens linearly at large quark separations. 
\end{abstract}
 
 \maketitle 
\section{Introduction} 
  Revealing the color field distribution in the nucleon is a subject of fundamental importance to quantum chromodynamics (QCD) and confinement. Lattice QCD simulations provide a first principle source of knowledge about how the energy distribution manifests itself among a system of three static quarks (3Q). This has to do with the relevant ansatz that accurately parametrizes and models the non-abelian force that binds the nucleon. The color distribution due to a 3Q system has been of a problem of reviving interest of lattice simulations and has been revisited with variety of lattice techniques~\cite{Bissey,Okiharu2004745,Bornyakov,Ichie}. Even though, an important aspect of this problem yet remains to be thoroughly investigated. That is, the energy distribution associated with the 3Q system at finite temperature. Tackling the problem of the gluonic distribution from this perspective involves the employment of a methodologically different set of unbiased hadronic operators. In addition to that, revealing the changes of the gluonic profile of the (3Q) system under various temperature conditions would certainly contribute to our perception of the underlying gluonic picture and the associated gluon dynamics. In fact, the distribution of gluonic fields in the baryon at high temperature, before quantum chromodynamics (QCD) undergoes a phase transition, is unknown in detail and has not yet been under the scrutiny of the lattice approach. 

   Most of our current understanding of the (3Q) confining force is based on the analysis at zero temperature ~\cite{Takahashi:2000te, Bali:2000gf, Alexandrou:2001ip, Takahashi:2002bw,Alexandrou:2002sn,K1,K2}. The parametrization which provides the best possible fits of the lattice data of the measured 3Q system potentials has been controversial for a long period of time~\cite{sommer,Thacker,Alexandrou:2002sn,Takahashi:2000te}. However, recent lattice QCD findings regarding the three quark potential are settled to support the so-called $\Delta$-ansatz parametrization for small quark separation distances of $ R < 0.7$ fm and the Y-ansatz for $0.7<R<1.5$ fm ~\cite{Alexandrou}. The $\Delta$ ansatz accounts for a confining potential built up as a sum of two-body forces, the string tension is half that in the corresponding $ Q \overline{Q}$ system and the confining part of the baryonic potential is in proportion to the perimeter of the triangle set up by the 3Q system. On the other hand, if the confining potential is proportional to the sum of the distances from the quarks to the Fermat point with a string tension the same as that in the $Q\bar{Q}$ system, then due to its shape, this potential is known as the Y-ansatz, giving rise to a three-body term relevant to a genuine interaction channel of the non-abelian force. 

  Ambiguities are known to arise, however, in the calculations of the gluonic distribution in the 3Q system at zero temperature. The energy distribution may be vulnerable to systematic errors associated with excited-state contamination~\cite{Okiharu2004745} when constructing the static baryon using a Wilson loop operator. The configuration of the spatial links that best minimize the potential has to be adopted before hand. Associated with this arbitrariness in tuning the ground state operator are the excited state potentials which manifest themselves in the revealed gluonic profiles as a bias reflecting the form of spatial links of the Wilson loop operator~\cite{Okiharu2004745}. The $L$ shape baryon operator provides a pronounced evidence where the flux distribution mimics the source~\cite{Bissey}. 

 The isolation of the ground state is challenging in the case of field-distribution calculations which involve four-point correlations rather than the ground state potential which is extracted in the large time limit of a three-point correlation~\cite{Okiharu2004745}. For example, Euclidean time evolution in the three quark Wilson loop operator results in observable broadening of the junction in the Y-shaped configuration~\cite{Bissey}. Statistical noise, nevertheless, imposes a practical constraint on any further increase in the exponentially decaying operator. 

  In this investigation, the static baryonic states are accounted for by means of Polyakov loops. This provides a gauge invariant operator which acquires a methodological importance~\cite{Bornyakov} due to the ability to construct an unbiased 3-quark operator without recoursing to a particular assumption regarding the form of the configuration of the spatial links in Wilson loops or the ultraviolet properties of these parallel transporters~\cite{Bakry:2011cn}. While carrying out energy density calculations into the zero temperature regime requires substantial numerical simulations with regard to the CPU time as well as the memory storage, the use of these stringless hadronic operators for revealing the energy distribution at finite temperature is still an attractive idea from the practical feasibility point of view. This can be studied in conjunction with the thermal effects.

  At finite temperature, pure Yang-Mills SU(3) lattice simulations for the action density in the mesonic sector display a flux distribution with a vibrating string-like shape. The density distribution shows a non-uniform pattern with an almost constant cross section in the intermediate distance region $0.5 \leq R \leq 1$ fm and non-constant cross section at larger quark separations~\cite{PhysRevD.82.094503}. The non-uniformity of the action density coincides with only a small decrease in the $Q \overline{Q} $ effective string tension $ \sigma $~\cite{Kac} suggesting the ground state may also display a non uniform action density distribution~\cite{Bakry:2011cn}. 

  In this paper, we generalize this analysis to the distribution of the color field inside the baryon. We consider one temperature near the end of the QCD plateau region at $T/T_{c} \approx 0.8$, and other just before the deconfinement point at $T/T_{c} \approx 0.9$. The three infinitely heavy quarks are accounted for by means of Polyakov loops of the same time orientation. The field strength inside the corresponding quark system is revealed by correlating an improved action density operator~\cite{Bilson} to these gauge-invariant hadronic operators. Gauge-field smoothing~\cite{Morningstar}, in addition to a high statistics gauge-independent~\cite{Bissey} averaging is employed to enhance the signal to the noise. This noise reduction approach can be employed in a  controlled and systematic manner that has been proved effective in keeping the physics intact in the case of the static meson~\cite{PhysRevD.82.094503}. The analysis on either the $Q\bar{Q}$ force or the action density shows that smearing leaves no effect on the corresponding measurements taken for quark source separation distance scales greater than the diameter of the Brownian motion of a diffused link i.e. the characteristic diameter of smearing. Moreover, the systematic effects associated with this UV filtering procedure on the gluonic profile has been reported in detail in Ref.~\cite{Bakry:2010sp}. 

  The analysis on the smearing effects is revisited in this work for the 3Q force. The relevant distance scale where the physics is preserved is established. After identifying this scale, the characteristics of the action density profile are presented for selected 3Q configurations and contrasted at the two considered temperatures. 
  
  The map of this paper is as follows: In Section~\ref{Measurements} and Section~\ref{Statistics}, the details of the simulations and noise reduction techniques are described. The force in the 3Q system for selected configurations is evaluated in Section~\ref{Forces}. In Section~\ref{ACT}, the main aspects of the gluonic profile of the baryonic action density is analyzed and contrasted at the two temperatures. In the last Section ~\ref{conc}, the conclusions are provided.

\section{\label{Measurements}Measurements} 
  The infinitely heavy quark state is constructed by means of Polyakov loop correlators. In the confinement phase, for pure SU(3) gauge configurations, the correlators respect the center symmetry transformation
\begin{equation}
   \tilde{U}_{\mu=4}(x,n_{\tau} = 1) = C \, U_{\mu=4}(x,n_{\tau} = 1),
\end{equation}

\noindent where center $C$ of the gauge group $SU(3)$ is all the elements $z$ such that $ zgz^{-1} = g $, with $ g\in SU(3) $ or $z=\exp(2 \pi i l/3 ) \in Z(3) $ with $ l=0,1,2 $. The form of the center symmetry preserving baryonic correlators is then
\begin{align*}
\langle \mathcal{P}_{3Q}(\vec{r}_{1}, \vec{r}_{2}, \vec{r}_{3})\rangle &\to \langle \tilde{P}(\vec{r}_{1} ) \tilde{P}(\vec{r}_{2} ) \tilde{P}(\vec{r}_{3} ) \rangle \notag  \\
& = \langle e^{2 i \pi l} P(\vec{r}_{1})P(\vec{r}_{2})P(\vec{r}_{3}) \rangle \\
&= \langle P(\vec{r}_{1})P(\vec{r}_{2})P(\vec{r}_{3}) \rangle~~~,
\end{align*}
 where the Polyakov loop is given by
\begin{equation}
   P(\vec{r}) = \frac{1}{3}\mbox{Tr} \left[ \prod^{N_{t}}_{n_{t=1}}U_{\mu=4}(\vec{r},n_{t}) \right],
\end{equation} 
which corresponds to three Polyakov lines all in  the same time direction.

\noindent After the construction of the gauge-invariant color-averaged quark states, subsequent measurement by a gauge-invariant action density operator $S(\vec{\rho},t)$ is taken at the spatial coordinate ${\vec{\rho}}$ of the three dimensional torus corresponding to each Euclidean time slice.  The action density operator is calculated via a highly-improved $\mathcal{O}(a^{4})$ three-loop improved lattice field strength tensor~\cite{Bilson}.

\noindent A scalar field that characterizes the gluonic field can be defined as

\begin{eqnarray}
\mathcal{C}(\vec{\rho},\vec{r}_{1},\vec{r}_{2},\vec{r}_{3}) & = \dfrac{\langle\mathcal{P}_{3Q}(\vec{r}_{1},\vec{r}_{2},\vec{r}_{3}) \, S(\vec{\rho})\rangle } {\langle \mathcal{P}_{3Q}(\vec{r}_{1},\vec{r}_{2},\vec{r}_{3})\rangle\, \langle S(
\vec{\rho}) \rangle}, 
\label{Cor}
\end{eqnarray}

\noindent for baryonic systems, where $< ...... >$ denotes averaging over configurations and lattice symmetries, the vectors $\vec{r}_{i}$ define the positions of the quarks and $\vec{\rho}$ the position of the flux probe. Cluster decomposition implies $C \longrightarrow 1$ away from the quarks. 

  In this investigation, we have taken $10,000$ measurements at temperature $T/T_{c} = 0.8 $, and $6,000$ measurements at temperature  $T/T_c = 0.9 $. The measurements are taken on hierarchically generated configurations. The gauge configurations are generated using the standard Wilson gauge action on lattices with a spatial volume of $36^{3}$. Gauge configurations are generated with a coupling value of $ \beta = 6.00$. The lattice spacing at this coupling is $a=0.1$ fm~\cite{PhysRevD.47.661}. After each 1000 of updating sweeps, $n_{\mathrm{sub}}=20$ or $12$ measurements separated by 70 sweeps of updates are taken for the two lattices corresponding to $T/T_c \approx 0.8$ and $T/T_{c} \approx 0.9 $ respectively. These sub measurements are binned together in evaluating Eq.~\eqref{Cor}. The total measurements are taken on 500 bins.    

  The gluonic gauge configurations are generated with a pseudo-heatbath algorithm~\cite{Cabibbo}. The heatbath is implemented by (FHKP)~\cite{Fabricius,Kennedy} updating on the corresponding three SU(2) subgroups. Each update step consists of one heatbath sweep and 4 micro-canonical reflections.

\section{\label{Statistics}Statistics}
  Gauge-independent noise reduction can be employed by making use of the space-time translational invariance of the hyper-toroid. By space-translational invariance, measurements of the action density operator is repeated 
for each translated Polyakov loops correlator. The measurements are then averaged over the spatial volume of the lattice. The periodicity in the time direction also allows averaging over the time direction   
\begin{eqnarray}
   S(\vec{\rho})=\frac{1}{N_{t}} \sum_{n_{t}=1}^{N_{t}}S(\vec{\rho},t).
\end{eqnarray}
  where $N_{t}$ is the number of time slices of the lattice. The symmetry of the quark positions can be also exploited to gain a reduction in the statistical uncertainties. The flux-distribution has been averaged around all the symmetry planes of a given quark configuration.

   Local action reduction by smearing the gauge links has been performed on the whole four dimensional lattice. This procedure can be applied for correlating operators with polyakov loops. For example, the correlations with the topological charge has been studied in Ref.~\cite{thurner} using the Cabbibo-Marinari cooling. Smearing the gauge field can be particularly helpful in reducing the statistical noise associated with evaluating the Polyakov loop correlators. However this step may result in the elimination of short distance physics and one has to be careful with regard to the number of smearing sweeps and the relevant distance scale where the physical observables are extracted. In the next section, we study the effects of gauge field smoothing on the 3Q force and determine the distance scale and corresponding smearing level where the physical observables are left intact. We smooth the gauge field with an over-improved stout-link smearing algorithm~\cite{Moran}. The scheme of the over improvement of this algorithm is such that a minimal effect on the topology of the gauge field~\cite{Moran} is ensured.  

 In the standard stout-link smearing~\cite{Morningstar}, all the links are simultaneously updated. Each sweep consists of a replacement of all the links by the smeared links,
\begin{equation}
  \tilde{U}_\mu(x) = \mathrm{exp}(i Q_\mu(x) ) \, U_\mu(x) \,,
  \label{eqn:stoutlinksmearedlink}
\end{equation}
with
\begin{align*}
  Q_\mu(x) & = \dfrac{i}{2}(\Omega_\mu^\dagger(x) - \Omega_\mu(x)) \notag\\
  & \quad - \dfrac{i}{6}\rm{tr}(\Omega_\mu^\dagger(x) - \Omega_\mu(x)) \,,
\end{align*}
and
\begin{align*}
  \Omega_\mu(x) = \left( \sum_{\scriptstyle  \nu \ne \mu}
  \rho_{\mu\nu} \Sigma_{\mu\nu}^\dagger(x) \right) U_\mu^\dagger(x)\, ,
\end{align*}
where $\Sigma_{\mu\nu}(x)$ denotes the sum of the two staples touching $U_\mu(x)$ which reside in the $\mu-\nu$ plane.
In the over-improved algorithm, however, the square staple is replaced by a combination of plaquette and rectangular staples. This ratio is tuned by the parameter $\epsilon$~\cite{Moran}. In the following we use a  value of $\epsilon = -0.25$, with $\rho_\mu = \rho = 0.06 $. 
 
  Despite the noise reduction at short distance quark separation, the signal drawn in statistical noise at large distances if one opt merely to the link integration method~\cite{Parisi1983418} in evaluating the Polyakov loops in Eq.~\eqref{Cor}. 

While the link integration method ~\cite{Parisi1983418} provides efficient noise reduction at small quark separations for the expectation values of Polyakov loops in Eq.~\eqref{Cor}, noise remains problematic at large separations. The method is expected to be efficient for the $Q \bar{Q}$ source separation range $R \leq 0.5$~\cite{1985PhLB15177D} but has to be supplemented with a large number of measurements for the calculations of the larger distance $Q\bar{Q}$ potential~\cite{Kac}. The number of measurements has to be increased significantly for the corresponding 3Q potential calculations with a three point correlator. The action density calculations would be, on the other hand, even more challenging in both cases~\cite{Bakry:2010sp}. The exponential reduction of noise provided by the leveling approach of the L\"uscher Weiss (LW) algorithm requires the hierarchical integration to be carried out over time sub-slices larger than the deconfinement temporal extent $t_{sub}>1/T_{c}$~\cite{Luscher:2001up}. For the temperatures considered here, the division of measurements into binned sub-measurements resembles a one level implementation of the LW method, with updating the last time slice.

\section{\label{Forces}Forces in the static baryon}  
  Unlike the force which is extracted from a three point correlator, the flux characterization Eq.~\eqref{Cor}, involves a four-point correlation function in the numerator presenting additional challenges with respect to the signal to noise level. The lattice space-time and configuration space symmetries can be auxiliary in enhancing the signal to noise ratio,  however, a four dimensional gauge smoothing has to be employed to obtain a signal. 

  The force in the 3Q system is a physical observable of direct relevance to the properties of the underlying energy distribution. In the following we consider the effects of the gauge smoothing procedure on the force experienced by a test color charge. 
  This can give indications on the relationship between source separation distance and the number of smearing sweeps where the changes in physics is minimal. Similar techniques have been adopted in Ref.~\cite{PhysRevLett.102.032004} in the determination of the large distance $Q \overline{Q}$ force in vacuum with different levels of hyperbolic (HYP) smearing~\cite{PhysRevD.82.094503}. In the following, we consider the evaluation of the force via three Polyakov loop correlators. 


\begin{figure}[!hpt]
\begin{center}
\includegraphics[width=7cm]{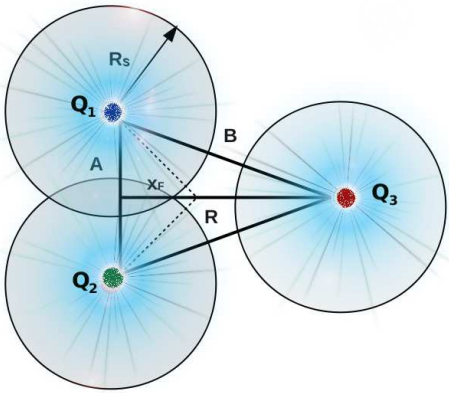}
\caption{\label{fuz} Schematic diagram for the isosceles configuration of the 3Q system. The large balls represent the motion of the diffused field of characteristic smearing radius of $R_{s}$ centered at the quarks (small spheres).}
\end{center}
\end{figure}


\begin{figure}[!hpt]
\begin{center}
\subfigure[$A=0.6$ fm]{\includegraphics[width=8cm]{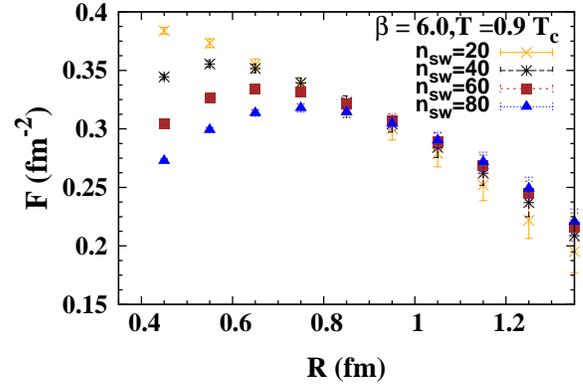}}
\subfigure[$A=0.8$ fm]{\includegraphics[width=8cm]{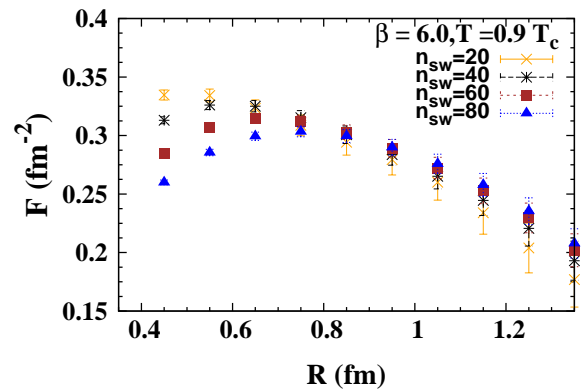}} 
\subfigure[$A=1.0$ fm]{\includegraphics[width=8cm]{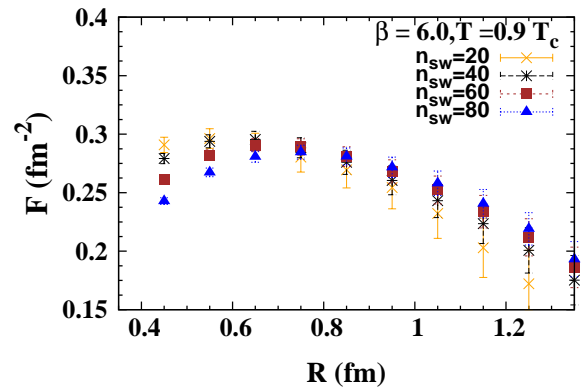}} 
\caption{\label{BF} The force for the isosceles 3Q configurations with base lengths (a) $A=0.6$ fm, (b) $A=0.8\,$ fm and (c) $A=1.0$ fm, respectively. The x-axis denotes the position $R$ of the third quark. Smearing effects are manifest for $R<0.95$ fm, $R<0.85$ fm and $R<0.75$ fm for $A=0.6$ fm, $A=0.8$ fm, and $A=1.0$ fm. Only subtle smearing effects remain beyond these distance scales.}
\end{center}
\end{figure}
  For several levels of smearing corresponding to $n_{sw}=\{20,40,60,80\}$, we numerically evaluate the force on a test color charge, assuming the transfer matrix interpretation is preserved as justified in Ref.~\cite{Bakry:2010sp}, the 3Q Potential can be identified via a three loop correlator as    

\begin{align*}
\langle \mathcal{P}_{3Q}\rangle &= \langle \mathcal{P}(\vec{r}_{1} ) \mathcal{P}(\vec{r}_{2} ) \mathcal{P}(\vec{r}_{3} ) \rangle \\
 &= \exp(-V_{3Q}(\vec{r}_{1},\vec{r}_{2},\vec{r}_{3})/T).
\end{align*}
 The force on the third quark $Q_3$ for the isosceles triangle configuration illustrated in Fig.~\ref{fuz} is measured through the definition of the derivative on the lattice~\cite{Sommer,luscher}  
\begin{align}
\label{force}
F_{Q_{3}}=&-\dfrac{\partial V (R;A)}{\partial R}|_{R+\frac{a}{2}}\notag \\
        =& \frac{1}{2\,a T}\log\left(\frac{\langle P(0,0)\,P(0,A) \,P(R,A/2) \rangle} {\langle P(0) \,P(0,A) \,P(R+1,A/2) \rangle}\right).
\end{align}

  The numerical values of the force measured on smeared configurations are reported in Fig.~\ref{BF} for three isosceles bases, $A=0.6,0.8\, \rm{and} \,1.0$ fm. The repeated measurements on the data sets corresponding to increasing smearing levels indicate, in general, invariance of the force experienced by the test charge $Q_3$ under smearing at large distances. The loss of short distance physics is pronounced at small values of $R$ which decreases as we increase the length of the isosceles base quark configuration. In the following, our consideration of different isosceles 3Q configurations enable a systematic identification of the distance scale beyond which a given level of smearing has little effect on the physical observables. 

  Define $R_F(n_{\rm{sw}})$ to be the minimal distance beyond which a smearing sweeps up to $n_{\rm{sw}}$ does not affect the force Eq.~\eqref{force}. The values of $R_F$ can be read from Fig.~\ref{BF}. Table~\ref{RtrustB} summarizes the values of $R_F$ for each isosceles configuration and smearing level. Since the effects of smearing relate also to the length of the isosceles base, we list for comparison the values of the corresponding effective range, $B_F$, defined as $B_{F}=\sqrt{R_{F}^2+A^{2}/4}$, and also the distance from a quark at the base of the triangle to the Fermat point of the triangle $L_{F}$ ( see Fig.~\ref{fuz}). 
\begin{table}
\caption{\label{RtrustB}The characteristic radii  $B_{F}=\sqrt{R_{F}^2+A^{2}/4}$ for the baryonic system at each smearing level for each configuration of Fig.~\ref{BF}.}
\begin{center}
\begin{tabular}{c|cc|cc|cccc}\hline
Config & $\,\,A= 0.6$&&$\,\,A=0.8$&&$\,\,A=1.0$&  \\
    & $\,\,L_F= 0.35$&&$\,\,L_F=0.46$&&$\,\,L_F=0.58$&  \\\hline
$n_{\rm{sw}}$ &$ R_{F}$ & $B_{F}$ &$R_{F}$&$ B_{F}$ & $R_{F}$ &$B_{F}$ \\\hline\hline
40 &0.65& 0.63 & 0.55 & 0.68     & 0.45&0.67     \\
60 &0.75& 0.80 & 0.65 & 0.76 & 0.55& 0.74\\
80 &0.85& 0.90     & 0.75 & 0.85 & 0.65& 0.82\\ \hline\hline
\end{tabular}
\end{center}
\end{table}

  Clearly the range $R_F$ is decreasing with the increase of the length of the isosceles base, $A$, indicating that smearing around the charges residing on the isosceles base decreases the force exerted on the color test charge $Q_3$ as the charges $Q_1$ and $Q_2$ become closer as in Fig.~\ref{fuz}. Inspection of the corresponding values of the above defined $B_F$, on the other hand, show that the decrease of $R_{F}$ with the increase of $A$ is such that the length of $B_{F}$ is approximately constant. To gain an insight to what these observations may imply, we study the characteristics of the Brownian motion of the diffused field, and also the analogous values of smearing threshold $R_F(n_{\rm{sw}})$ measured for the $Q \bar{Q}$ system.

  The diffuse field is Gaussian distributed~\cite{Takahashi:2002bw} through a ball centered at position ${\bf r}$ whose evolution with a smearing time $\tau$, in a four-dimensional smearing scheme~\cite{PhysRevD.82.094503} is given by  
\begin{equation}
	G({\bf r};\tau)=
	{1\over (4\pi D  \tau)^{3\over 2}}
	\exp\left[ - {{\bf r.r} \over 4 D  \tau} \right],
\end{equation}
with $D$ describing the diffuseness of the field. The diffused field characteristic radius is defined as
\begin{align}
R_{s} \equiv& \left(\frac{\int d^{3}{\bf r} \, G({\bf r};\tau)   {\bf r}^{2} }
{\int d^{3}{\bf r} \, G({\bf r};\tau)}\right)^{1/2}. \notag \\
=& a \sqrt{\rho \,c \, n_{sw} }.
\label{size}
\end{align}

\noindent The proportionality constant $c$ scales the number of smearing sweeps $n_{sw}$ in the improved stout-link smearing algorithm defined above with respect to APE smearing as defined for instance in Refs.~\cite{Bissey,Bakry:2011cn}. The calibration proceeds via comparing the respective number of smearing sweeps in each smearing scheme with respect to a given threshold~\cite{Bonnet} (the reconstructed action-density~\cite{Bilson} normalized to a single instanton action $S/S_{0}$). This yields a value of $c=6.15(3)$~\cite{PhysRevD.82.094503}. With $\rho=0.06$, the number of smearing sweeps in the improved stout-link smearing algorithm scales as half the number of the corresponding smearing sweeps in APE smearing with the smearing parameter $\alpha=0.7$. 
    
  After identifying this characteristic smearing range, the values of $R_s(n_{\rm{sw}})$ are compared to the corresponding values of $R_{F}(n_{\rm{sw}})$ for the $Q\bar{Q}$ system ~\cite{PhysRevD.82.094503} in Table~\ref{RtrustM}. Inspection of the values reported for the mesonic force unveils that $R_F$ is roughly equivalent to twice the smearing radius $R_s$. This suggests that the mesonic force is invariant under the smearing operation applied on the whole four dimensional lattice as long as the fuzzed balls centered at the quark source links are non-overlapping. Similar analysis on the action density shows that the region free of smearing effects obeys the same invariance criterion~\cite{PhysRevD.82.094503}.

  The distance $B_F$ describes the minimal distance from the quarks $Q_{1,2}$ to $Q_3$ for which the measured force is invariant under a given number of smearing sweeps. The values of $B_{F}$ in Table~\ref{RtrustB} compare favorably with the values calculated for the quark--antiquark system in Table~\ref{RtrustM}. This indicates that the smearing effects are immaterial as long as the length of the isosceles side is such that the fuzzed balls around any of the color charges $Q_{1,2}$ and that around the test charge $Q_3$ are non-overlapping.  

\begin{table}[!hpt]
\caption{\label{RtrustM} The characteristic radii $R_{s}$ and $R_{F}$ at each smearing level for mesonic systems beyond which the force is unaffected by smearing~\cite{PhysRevD.82.094503}.}
\begin{center}

\begin{tabular}{cccc}\hline
 No.Sweeps &$ R_{F}$(fm) & $R_{s}$(fm) &$2\,R_{s}$(fm) \\\hline\hline
40 & 0.65 & 0.38&0.76 \\
60 & 0.75 & 0.47&0.94 \\
80 & 0.95 & 0.54&1.04\\ \hline
\end{tabular}
\end{center}
\end{table}

  For the smearing radii considered here, a slight overlap of the fuzzed ball around each quark $Q_{1,2}$ on the base of the isosceles is seen to have no observable effect on the force experienced by the test charge $Q_3$. This observation does not exclude the possibility of the three body channel of the interaction in the (3Q) system. The locus of the center of interaction may still be outside the two overlapping spheres in the base of the triangle. 

  In Table~\ref{RtrustB}, the distance from the quarks $Q_{1,2}$ to the Fermat point is indicated for each configuration. Simple variation calculus shows that for an isosceles triangle, the position of the Fermat point does not depend on the height of the triangle, $R$, and the locus is fixed merely by the length of the base of the triangle such that, $x_{F}=A/(2\sqrt{3})$. 


  In summary, for the quark position geometry considered in this work, the above analysis on the measured values of the force among a system of three quarks for each smeared data set of configurations suggest a conservative distance scale $\gamma$ beyond which the confining force on a given source is unchanged to be    

\begin{equation}
\gamma=2\,R_{s}.
\label{range}
\end{equation} 
  This restricts the number of smearing sweeps to be such that the characteristic diameter of smearing does not exceed the distance between at most two quarks. To take into account the distance between other sources, an additional conservative measure will be to keep the distance to the Fermat point from any of the quarks of a given configuration outside the radius of smearing. However, in the present case this may be immaterial since the differences in the force measurements are well within the statistical errors. Equation~\eqref{range} indicates the distance scales where a specific characteristic of the action density might be affected by gauge field smoothing. The corresponding effects of gauge field smoothing on the revealed gluonic profile will be discussed also on several occasions below.

\section{\label{ACT}Action Density}
\subsection{\label{profile}Flux Iso-surface Profile}

\begin{figure}[!hpt]
\begin{center}
\subfigure[$R=0.6$ fm, $n_{\rm{sw}}=60$.]{\includegraphics[height=5cm, width=5cm]{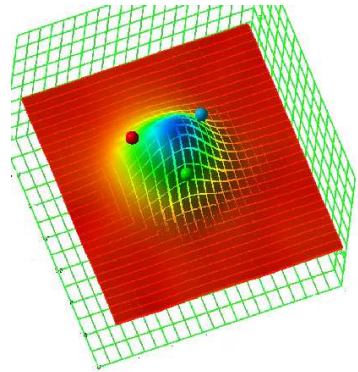}} \\
\subfigure[$R=0.8$ fm, $n_{\rm{sw}}=60$.]{\includegraphics[height=5cm, width=5cm]{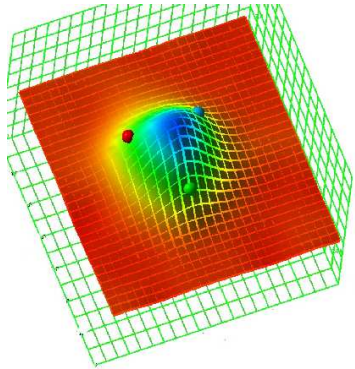} }\hspace{1cm}
\subfigure[$R=1.0$ fm, $n_{\rm{sw}}=80$.]{\includegraphics[height=5cm, width=5cm]{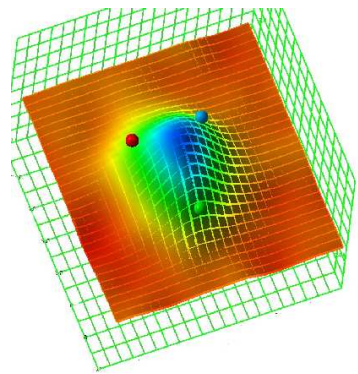}}
\caption{ \label{surf} Surface plot (Inverted) of the flux distribution $\mathcal{C}(\vec{\rho})$ of Eq.~\eqref{Cor} evaluated in the plane of the (3Q) system $\vec{\rho}(x,y,0)$, for isosceles configuration of base length $A=0.4$ fm and  separation distances (a)~$R=0.6~$ fm,~ (b)$~R=0.8$ fm and (c)~$R=1.0$ fm, at $\rm{T}=~0.8 ~\rm{T}_{C}$. The spheres refer to the positions of the quarks.}
\end{center}
\end{figure} 

\begin{figure}
\begin{center}
\includegraphics[width=7cm]{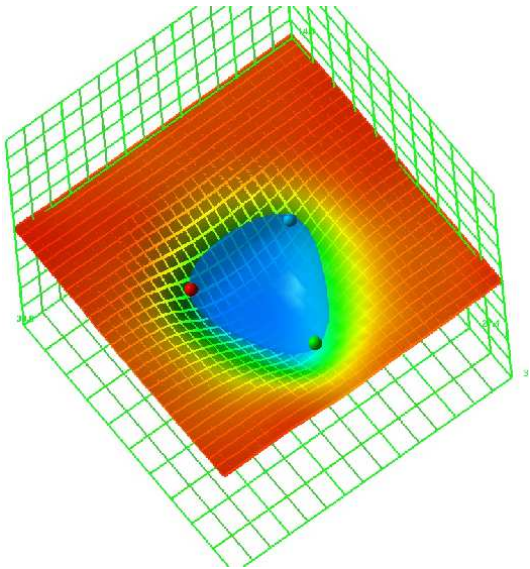}
\caption{\label{isosurf}The flux action iso-surface at the quark positions, plotted together with a surface plot for the density distribution $\mathcal{C}(\vec{\rho})$, in the 3Q plane at temperature $\rm{T}\,=\,0.9\,\rm{T}_{c}$, for equilateral triangular configuration $ R=1.1$ fm and $A = 1.0$ fm.}

\end{center}
\end{figure}

\begin{figure}
\begin{center}
\subfigure[$R=0.4$ fm, $n_{\rm{sw}}=40$.] {\includegraphics[width=4cm,angle=0]{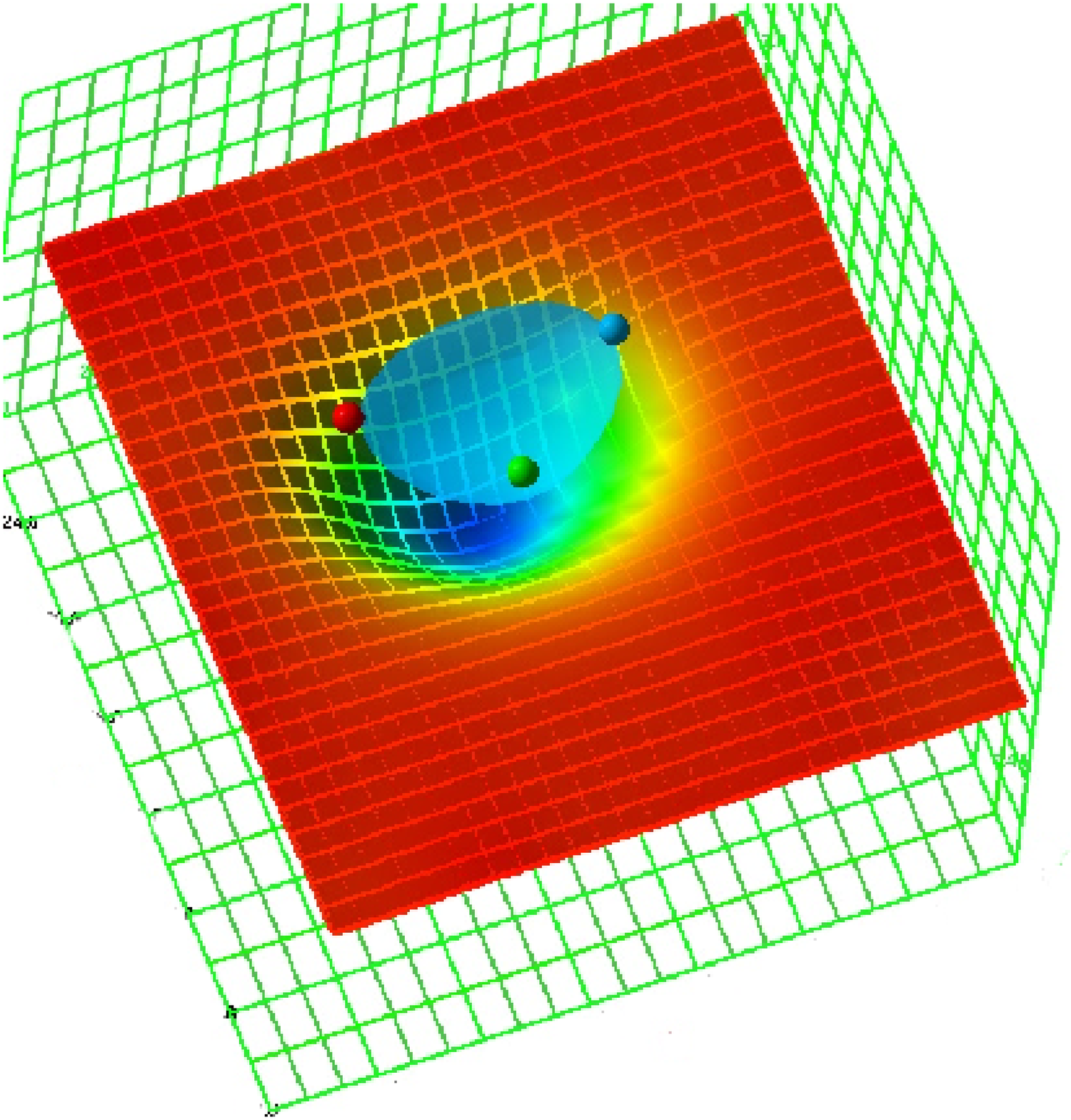} } 
\subfigure[$R=0.5$ fm, $n_{\rm{sw}}=40$.] {\includegraphics[width=4cm,angle=0]{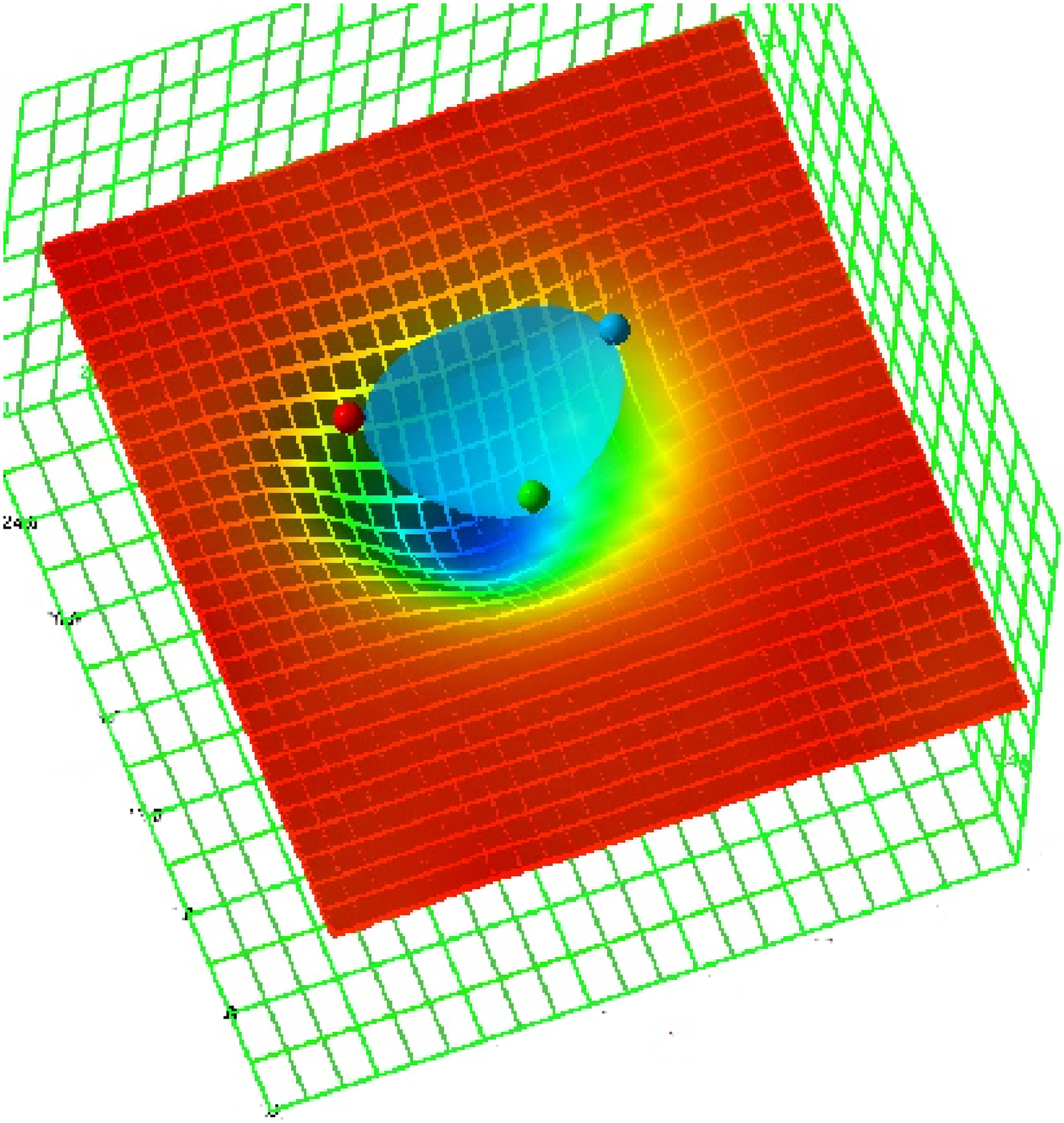} }  
\subfigure[$R=0.6$ fm, $n_{\rm{sw}}=60$.] {\includegraphics[width=4cm,angle=0]{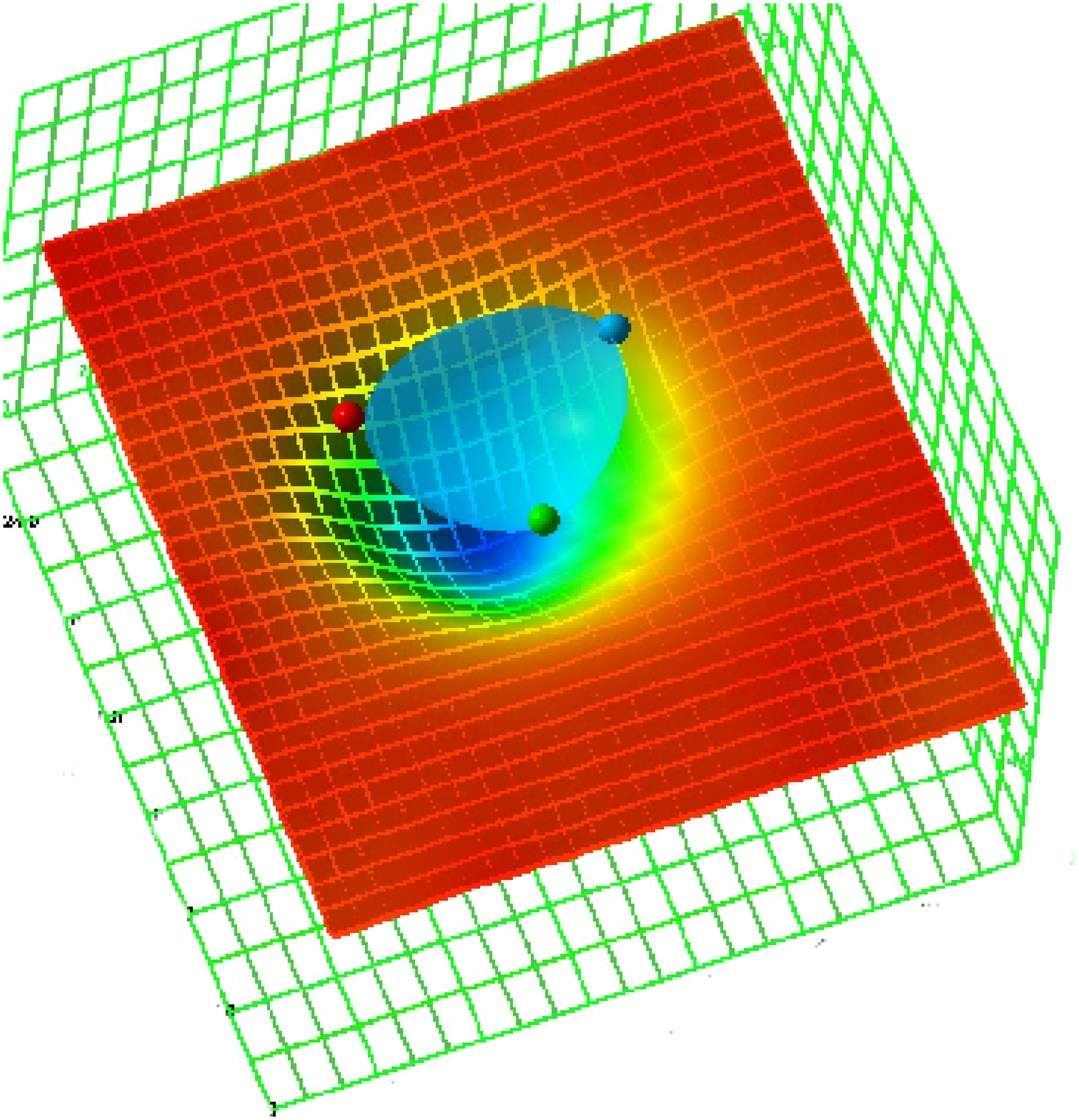} }  
\subfigure[$R=0.7$ fm, $n_{\rm{sw}}=60$.] {\includegraphics[width=4cm,angle=0]{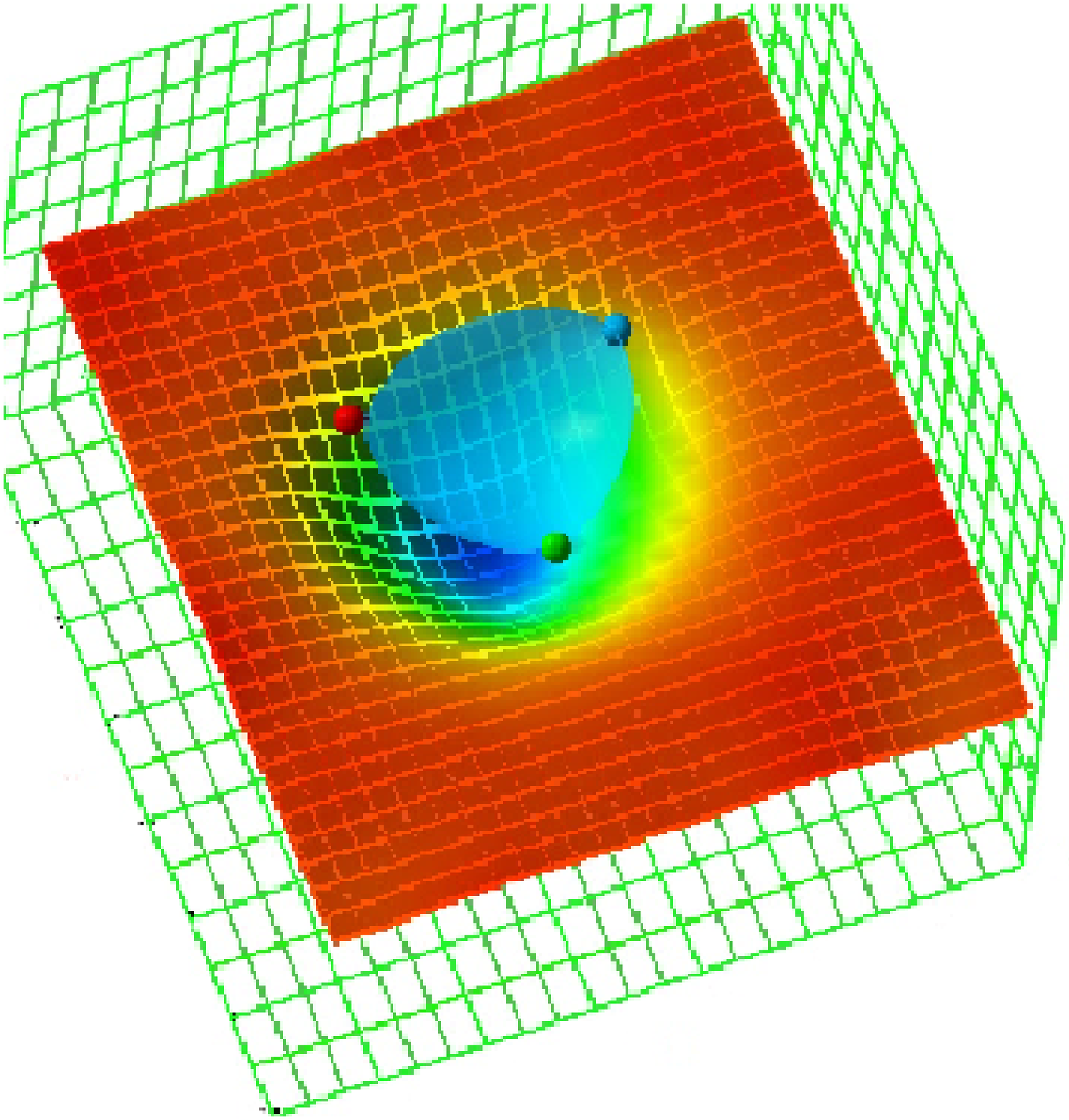} }  
\subfigure[$R=0.8$ fm, $n_{\rm{sw}}=60$.] {\includegraphics[width=4cm,angle=0]{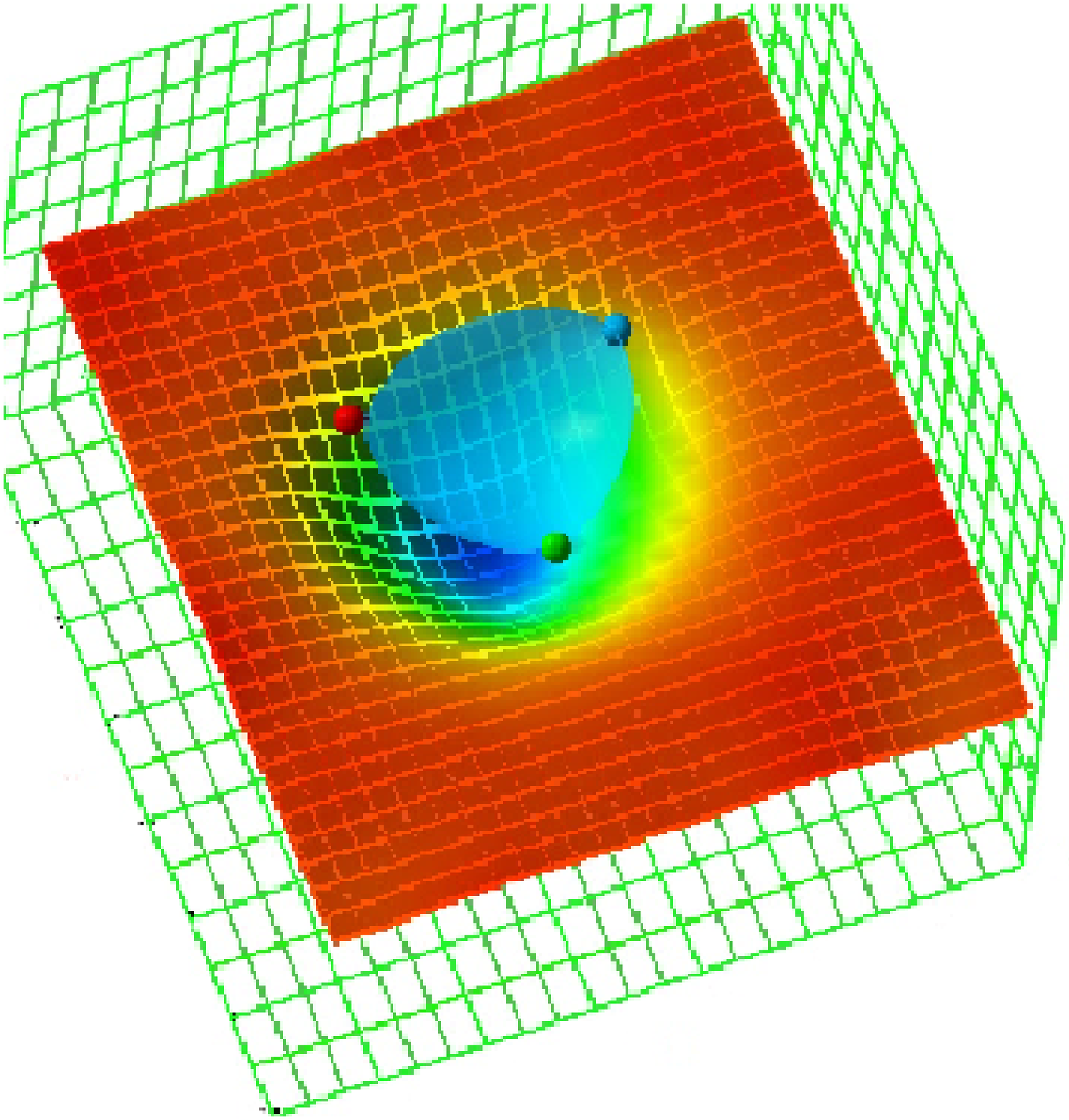} }  
\subfigure[$R=0.9$ fm, $n_{\rm{sw}}=80$.] {\includegraphics[width=4cm,angle=0]{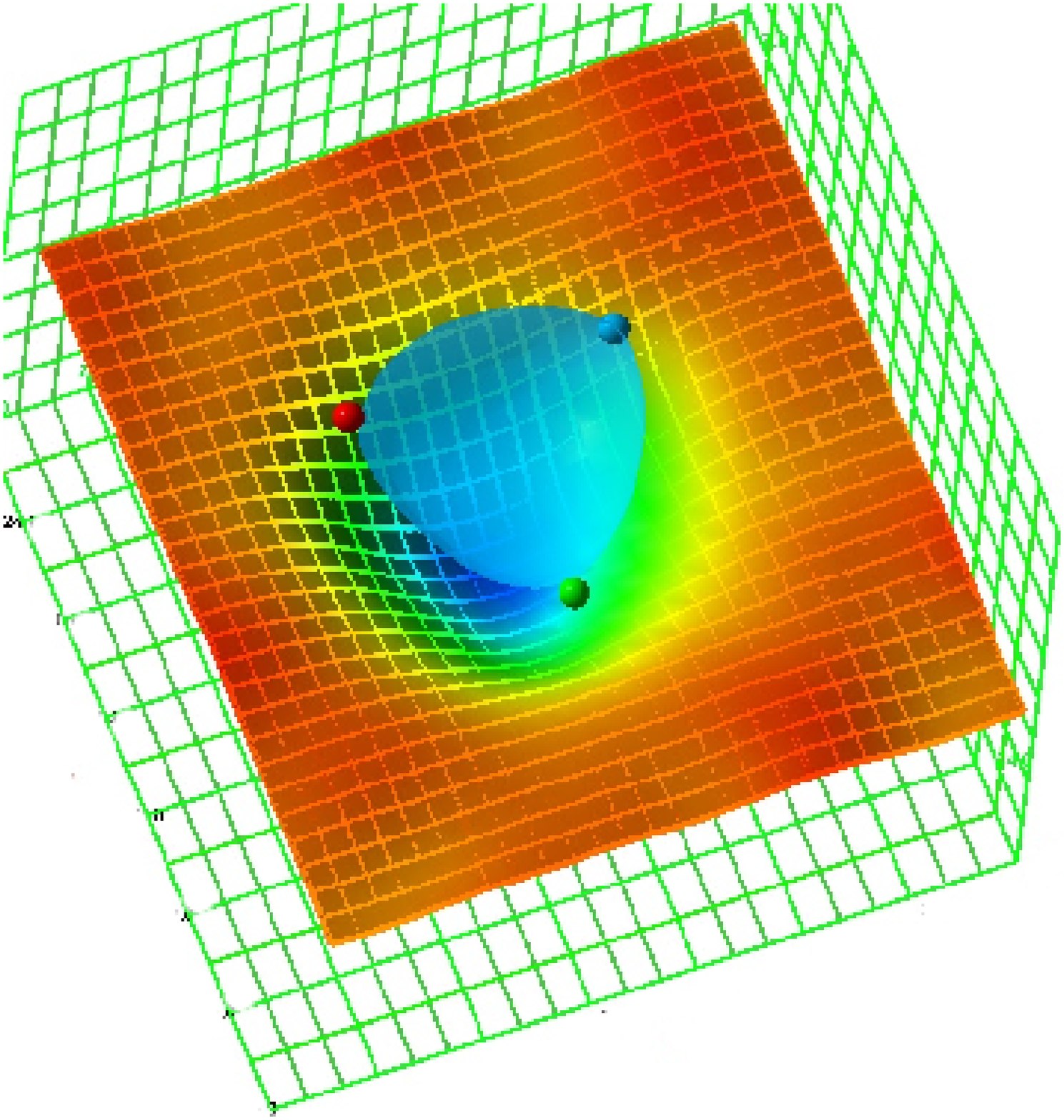} }  
\subfigure[$R=1.0$ fm, $n_{\rm{sw}}=80$] {\includegraphics[width=4cm,angle=0]{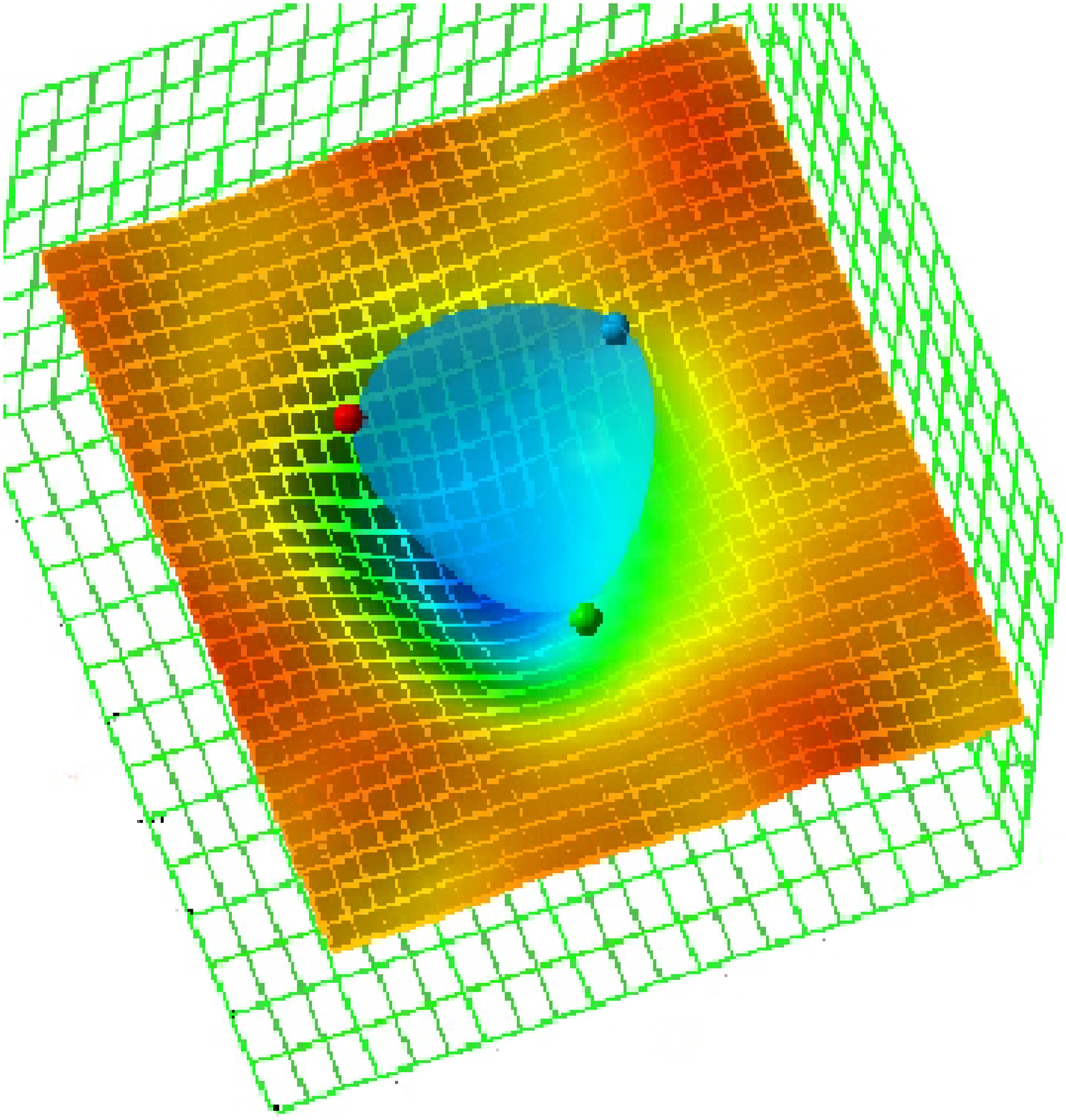}}  
\subfigure[$R=1.1$ fm, $n_{\rm{sw}}=80$] {\includegraphics[width=4cm,angle=0]{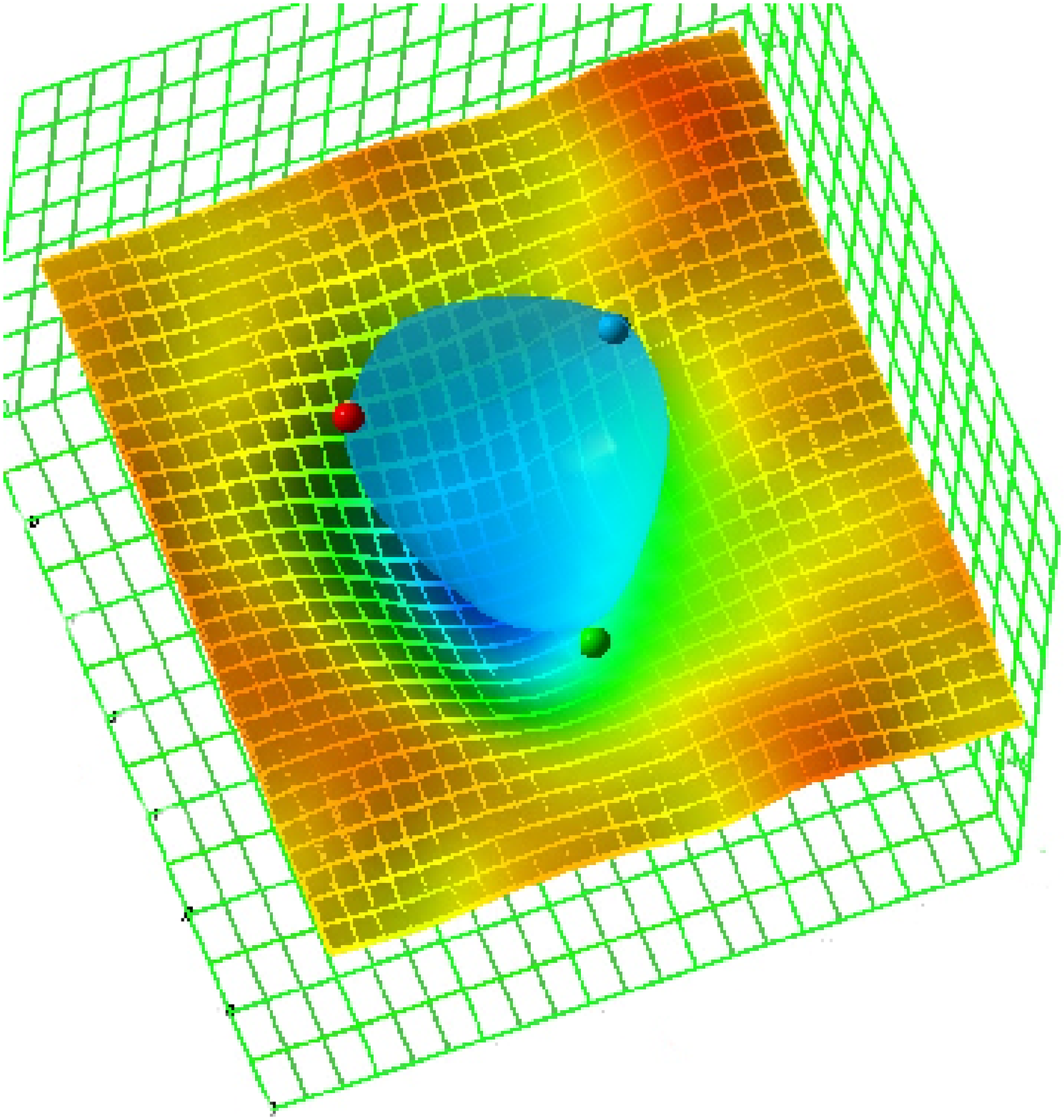}} 
\caption{\label{s}Surface plot in the plane of the 3Q system $\vec{\rho}(x,y,z=0)$ and iso-surface of the flux distribution $\mathcal{C}(\vec{\rho};\vec{r}_{1},\vec{r}_{2},\vec{r}_{3})$ for the isosceles configuration with $A=1$ fm and the third quark separation distance $R$ as indicated. $\rm{T}=~0.8 ~\rm{T}_{C}$.}
\end{center}
\end{figure}

\begin{figure*}
\begin{center}
\subfigure[$T=0.8\,T_c$.]{\includegraphics[width=7cm]{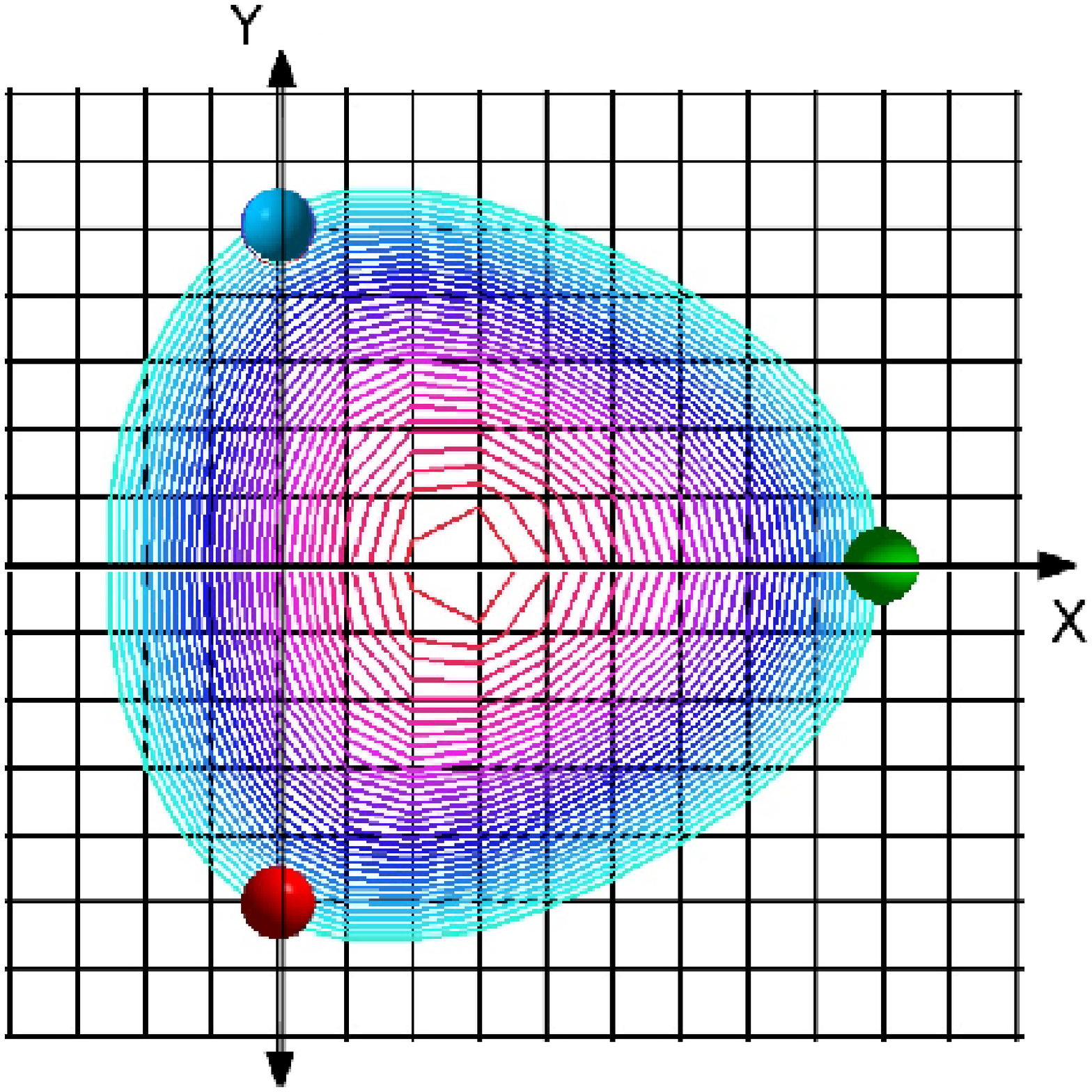}}
\subfigure[$T=0.9\,T_c$.]{\includegraphics[width=7cm]{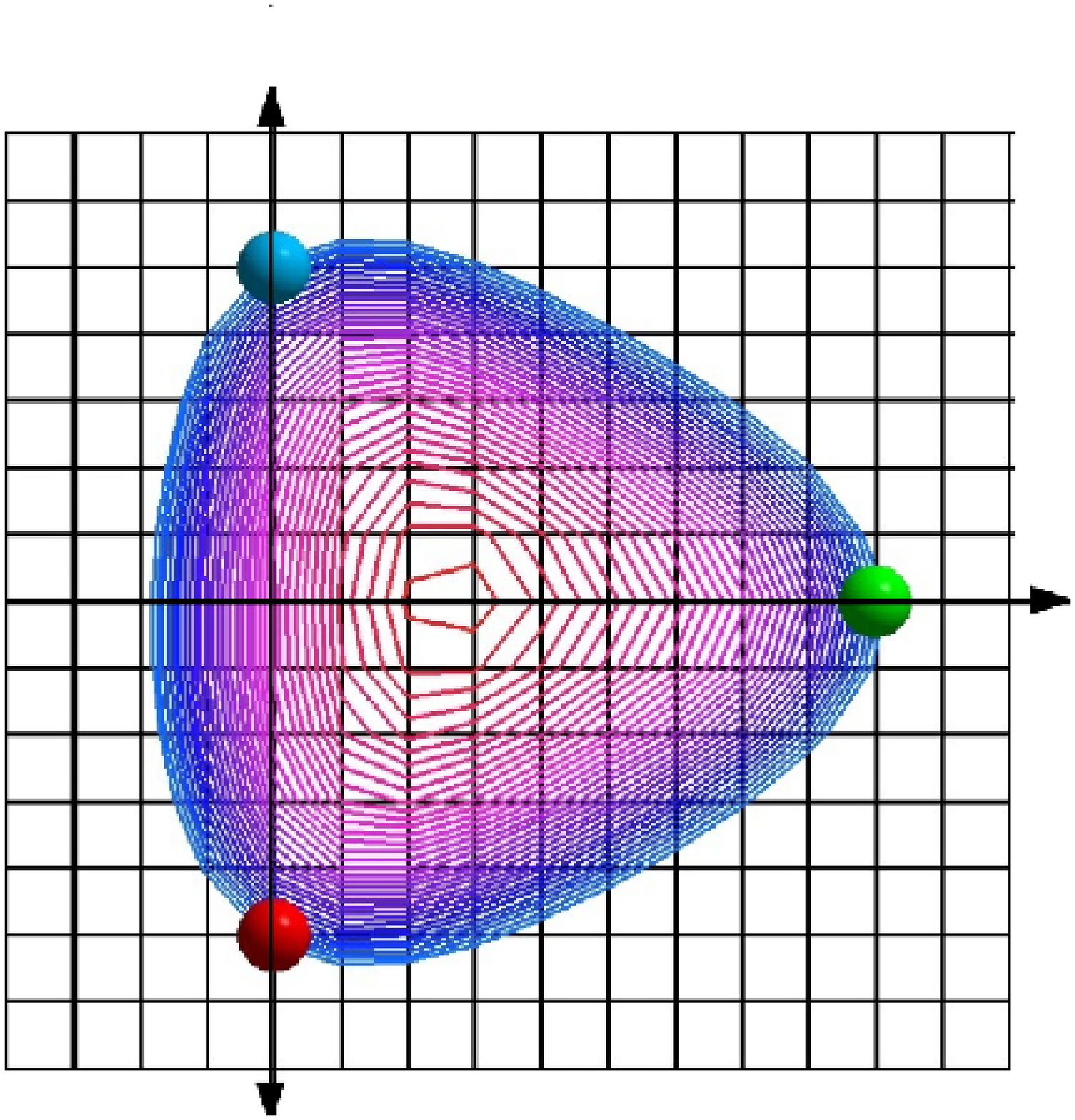}}
\caption{\label{contor} Compares the flux contour lines of the density distribution in the 3Q plane $\mathcal{C}$ for triangular base $A=1.0$ fm and third quark separation $R=0.9$ fm at (a)$T=0.8\,T_c$ and (b) $T=0.9\,T_c$, in $z=0$ plane.}
\end{center}
\end{figure*}

 The flux strength is measured as the correlation between the vacuum action-density, $S(\vec{\rho},t)$, and a gauge-invariant operator representing the quark states as provided by Eq.~\eqref{Cor}. The action density operator 
\begin{align}
    S(\vec{\rho})=\beta \sum_{\mu > \nu}\, \frac{1}{2} \mathrm{Tr}(F^{Imp}_{\mu \nu})^{2},
\end{align} 
\noindent is calculated through the $\mathcal{O}(a^{4})$ improved lattice version of the continuum field-strength tensor~\cite{Bilson}
\begin{align}
 F^{\mathrm{Imp}}_{\mu \nu} = \sum_{i=1}^{3} w_{i}\,C_{\mu \nu}^{(i,i)}. 
\label{clover}
\end{align}
where $C^{(i,i)}_{\nu\mu}$ are $i\times i$ link products in the $\nu \mu$ plane and $w_{i}$ are coefficients selected to remove $\mathcal{O}(a^{2})$ and $\mathcal{O}(a^{4})$ errors. The evaluation of this operator on smeared configurations filters out the UV divergences around the quark positions. The origin of the co-ordinate system is placed at the middle between the two quarks $Q_{2,3}$ on the $y$ axis at positions $\vec{\rho}(0,\pm \frac A 2,0)$ and at distance $R$ from the third quark, $Q_{3}$, at $\vec{\rho}(x=R,0,0)$. The quarks reside on the plane $\vec{\rho}(x,y,0)$. 

 On calculating Eq.~\eqref{Cor}, we find $\mathcal{C}(\vec{\rho})<1$, and $\mathcal{C}\simeq 1$ away from the quark positions. The density distribution in the plane of the quarks is plotted in Fig.~\ref{surf}.  In general, the action density distribution is non-uniformly distributed as revealed in Fig.~\ref{surf} through~\ref{s}. The distribution  $\mathcal{C}(\vec{\rho}(x,y,z=0))$ has an action density maximal curve along the middle line $\vec{\rho}(x,y=0,z=0)$ between the two quarks $Q_{1,2}$. With the increase of source $Q_3$ separation, the peak point along the maximal curve $\mathcal{C}(\vec{\rho}(x,y=0,z=0))$ shows only subtle movement, remaining near the Fermat point of the triangle. These results contrast the Wilson loop results at large separations~\cite{Bissey} where the action density assumes a constant amplitude along each arm of the Y-shaped profile. A convex curvature in the contour plot of flux density is manifest in Fig.~\ref{isosurf}(b). This also contrasts the density plots obtained using the Wilson loop where the flux density assumes a concave curvature.     

  Figure~\ref{isosurf} discloses the flux surface plot of $\mathcal{C}$ in the 3Q plane and associated iso-surface for an isosceles configuration corresponding to a base $A=1.0$ fm at the temperature $T/T_{c}=0.9$. The flux iso-surface displays a clear filled $\Delta$ shape distribution. By moving the third quark further away, i.e. by increasing $R$, the $\Delta$ shape is found to persist. The sequence of frames in Fig.~\ref{s} displays similar results, this time at $T=0.8\,T_c$. It is important to note that this geometrical form of the density plot manifests itself at a temperature near the end of the plateau of the QCD-phase diagram~\cite{Doi2005559} where the string tension has been reported to decrease only by a value around $10\%$~\cite{Kac}.  

The contour lines and iso-surface of the flux do not exhibit a significant change with the temperature scale. Similarly, the effects of smearing do not cause deformations of the iso-surface profile outside of smoothing the interpolation of the flux-lines. This has been observed ~\cite{ws} at relatively short distances employing the link integration for evaluating Polyakov lines in the flux strength characterization ~\eqref{Cor}. Even though the number of smearing sweeps selected for each graph are larger than the limits set by the invariance of the force analysis of the last sections, the rendered graphs are not sensitive to increased levels of smearing.   
\begin{figure}[!hpt]
\begin{center}
\includegraphics[width=8.5cm]{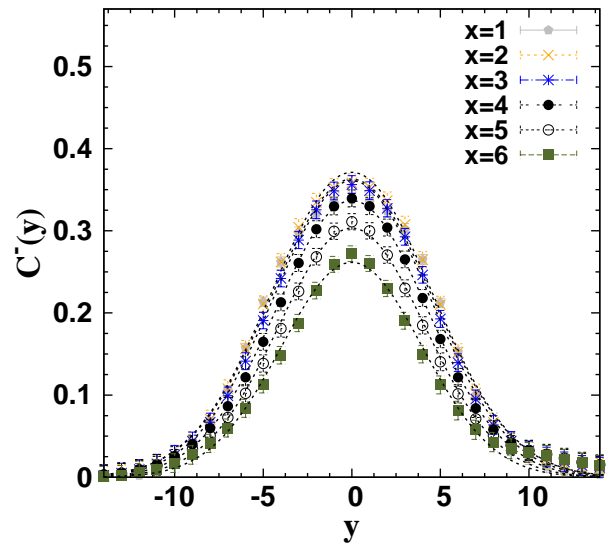} 
\caption{\label{gaussb} The density distribution $\mathcal{C'}(\vec{\rho})$ for the isosceles configuration with the base, $A\,=\,1.0$ fm, and height $R=0.8$ fm at $T/T_c=\,0.8$ ($n_{\rm{sw}}=60$ sweeps). Data are plotted for the transverse planes $x=1$ to $x=6$. The lines correspond to the Gaussian fits to the density in each plane $\vec{\rho}(x_{i},y,0)$. The highest amplitude lies close to the Fermat point plane $x=2.88$ of this 3Q configuration.} 
\end{center}
\end{figure}

  The flux distribution acquires a non-trivial transverse structure along the lines perpendicular to the $x$-axis. The fits of the transverse distribution along the lines $ \vec{\rho}(x_i,y,0)$ is returning good $\chi^{2}$ for a Gaussian distribution with varying amplitudes and widths from the third quark $Q_{3}$ position, $R$, to the $y$-axis as shown for instance in Figs.~\ref{surf} or~\ref{contor}. This symmetry about the $y$-axis in the $x-y$ plane also exists in the perpendicular $z$ direction.  

  In a mesonic system, the width of the flux distribution is cylindrically symmetric around the line joining the two quarks. However, the existence of a third quark away from the $y$-axis breaks the symmetry of the width profile and the measured widths perpendicular to the plane of the quarks do indeed differ from the widths in the quark plane. In the forth coming section, we focus on dissecting the profile properties of the flux distribution within the quark plane, while the asymmetry aspect ratios are reported separately in the last section.   

\begin{table*}
\caption{\label{1} The amplitude, $H_y(x_i)$ (scaled by a factor of $10^{1}$), of the flux distribution at each consecutive transverse plane $x_{i}$ from the quarks forming the base, $A$, of isoscoles triangle. The measurements for base source separation distance $A=0.6, 0.8$ and $1.0$ fm for the temperature $T/T_c = 0.8$ are indicated as a function of the third quark position, $Q_{3}$. }
\begin{center}
\begin{tabular}{cccccccccccccccc}\hline
  Plane& $x=1$&$x=2$&$x=3$ & $x=4$ &$x=5$&$x=6$&$x=7$&$x=8$& $x=9$ & $x=10$ & $x=12$  & $x=13$ \\ 
  $Q_{3}=R/a$  \\ \hline\hline
A=0.6 fm\\

07&4.12(2)&4.34(1)&4.28(1)&3.97(2)&3.48(2)&2.84(2)\\
08&4.24(2)&4.53(2)&4.56(2)&4.36(3)&4.00(3)&3.49(3)&2.85(3)\\
09&4.31(2)&4.63(3)&4.73(3)&4.63(4)&4.37(4)&3.99(4)&3.49(4)&2.85(3)\\
10&4.35(3)&4.67(5)&4.82(5)&4.79(6)&4.62(6)&4.33(5)&3.95(4)&3.46(3)&2.83(2)\\
11&4.36(5)&4.66(7)&4.84(7)&4.85(8)&4.74(8)&4.53(7)&4.24(5)&3.86(4)&3.39(3)&2.80(3)\\
12&4.35(6)&4.6(1)&4.7(1)&4.8(1)&4.8(1)&4.59(9)&4.35(7)&4.06(4)&3.72(4)&3.29(3)&2.73(3)\\
13&4.30(7)&4.4(1)&4.6(1)&4.7(1)&4.7(1)&4.5(1)&4.3(1)&4.06(6)&3.81(4)&3.53(4)&3.14(4)&2.64(3)\\
A=0.8 fm\\

07&4.71(5)&4.92(3)&4.80(1)&4.40(2)&3.80(3)&3.07(3)\\
08&4.79(5)&5.08(3)&5.06(3)&4.78(3)&4.31(4)&3.71(4)&3.01(3)\\
09&4.83(5)&5.14(4)&5.22(4)&5.03(4)&4.68(5)&4.22(4)&3.66(4)&2.98(3)\\
10&4.84(5)&5.13(5)&5.27(5)&5.18(5)&4.93(6)&4.58(5)&4.13(4)&3.60(4)&2.95(4)\\
11&4.84(5)&5.06(5)&5.24(5)&5.23(6)&5.08(6)&4.80(5)&4.44(4)&4.02(5)&3.51(6)&2.88(6)\\
12&4.84(5)&4.92(5)&5.12(5)&5.18(6)&5.12(6)&4.92(5)&4.61(3)&4.23(6)&3.8(1)&3.4(1)&2.75(9)\\
13&4.84(5)&4.70(4)&4.88(3)&5.00(3)&5.04(5)&4.92(5)&4.64(2)&4.27(8)&3.9(1)&3.5(1)&3.1(1)&2.6(1)\\
A=1.0 fm \\

07&5.17(9)&5.43(6)&5.33(3)&4.86(1)&4.15(2)&3.31(2)\\
08&5.22(9)&5.56(7)&5.58(5)&5.25(2)&4.69(1)&3.99(2)&3.19(3)\\
09&5.2(1)&5.58(9)&5.72(7)&5.53(5)&5.11(3)&4.55(2)&3.89(4)&3.14(4)\\
10&5.2(1)&5.5(1)&5.8(1)&5.71(9)&5.43(7)&4.99(6)&4.45(6)&3.82(7)&3.08(7)\\
11&5.26(1)&5.3(1)&5.7(1)&5.8(1)&5.7(1)&5.3(1)&4.9(1)&4.3(1)&3.7(1)&3.0(1)\\
12&5.4(1)&5.1(1)&5.5(1)&5.7(1)&5.8(1)&5.6(1)&5.1(1)&4.6(1)&4.0(1)&3.4(1)&2.7(1)\\
13&5.7(1)&4.7(1)&5.0(1)&5.4(1)&5.6(1)&5.7(1)&5.3(1)&4.6(1)&4.0(1)&3.4(1)&2.9(1)&2.4(1)\\\hline

\end{tabular}

\end{center}
\end{table*}

\begin{table*}
\caption{\label{2} The squared width, $r_y^{2}$, in lattice units, of the flux distribution as in Table~\ref{1}.}
\begin{center}
\begin{tabular}{cccccccccccccccc}\hline
 plane& $x=1$&$x=2$&$x=3$ & $x=4$ &$x=5$&$x=6$&$x=7$&$x=8$& $x=9$ & $x=10$ & $x=12$  & $x=13$  \\ 
  $Q_{3}=R/a$  \\ \hline\hline

A=0.6 fm\\
07& 9.7(1)& 9.6(0)& 9.3(0)& 9.0(1)& 8.6(1)& 8.2(1)\\
08&10.0(1)& 9.9(1)& 9.7(1)& 9.5(1)& 9.2(1)& 8.7(1)& 8.2(1)\\
09&10.2(1)&10.3(1)&10.2(1)&10.0(1)& 9.7(2)& 9.3(2)& 8.8(2)& 8.2(1)\\
10&10.6(1)&10.8(2)&10.6(2)&10.5(2)&10.3(2)& 9.9(2)& 9.4(2)& 8.8(2)& 8.0(1)\\
11&10.9(2)&11.5(3)&11.2(3)&11.1(3)&10.9(3)&10.6(3)&10.2(2)& 9.5(2)& 8.6(1)& 7.8(1)\\
12&11.3(2)&12.5(5)&12.1(4)&11.8(4)&11.6(4)&11.3(4)&11.0(3)&10.3(2)& 9.4(2)& 8.4(2)& 7.7(1)\\
13&11.7(3)&13.9(7)&13.2(6)&12.7(6)&12.4(6)&12.3(6)&12.1(5)&11.5(3)&10.4(2)& 9.2(2)& 8.3(2)& 7.9(2)\\

A=0.8 fm \\
07&12.2(2)&11.9(1)&11.3(0)&10.8(1)&10.2(1)& 9.5(2)\\
08&12.4(2)&12.2(1)&11.7(1)&11.2(1)&10.6(1)&10.0(2)& 9.2(2)\\
09&12.8(2)&12.6(1)&12.1(1)&11.6(1)&11.1(2)&10.5(2)& 9.7(2)& 8.8(2)\\
10&13.2(2)&13.1(2)&12.5(2)&11.9(2)&11.4(2)&10.9(2)&10.2(2)& 9.3(2)& 8.2(2)\\
11&13.6(2)&13.6(2)&12.8(2)&12.1(2)&11.5(2)&11.0(2)&10.4(1)& 9.6(2)& 8.5(3)& 7.5(3)\\
12&14.1(2)&14.0(2)&12.9(2)&12.1(2)&11.3(2)&10.8(2)&10.3(1)& 9.6(2)& 8.6(4)& 7.6(4)& 6.8(4)\\
13&14.8(2)&14.2(2)&12.7(1)&11.4(1)&10.6(2)&10.1(1)& 9.8(1)& 9.3(3)& 8.5(5)& 7.6(6)& 6.9(6)& 6.4(6)\\

A=1.0 fm \\
07&15.4(4)&14.6(3)&13.7(1)&12.9(0)&12.2(1)&11.4(2)\\
08&15.7(4)&14.8(3)&13.8(2)&13.0(1)&12.3(0)&11.5(1)&10.6(2)\\
09&16.1(5)&15.0(4)&13.8(3)&12.9(2)&12.1(1)&11.5(1)&10.6(2)& 9.5(2)\\
10&16.6(5)&15.1(5)&13.7(4)&12.6(3)&11.7(2)&11.0(2)&10.4(2)& 9.4(3)& 8.1(4)\\
11&17.0(6)&14.9(6)&13.3(6)&12.0(5)&10.9(4)&10.2(4)& 9.6(5)& 9.0(6)& 7.9(6)& 6.6(6)\\
12&17.2(7)&13.9(5)&12.2(5)&10.8(5)& 9.7(7)& 8.9(7)& 8.3(8)& 7.8(9)& 7(1)& 6(1)& 5(1)\\
13&17.1(9)&12(1 )&10(1)& 9(1)& 8(1)& 7(1)& 6(1)& 6(1)& 6(1)& 5(1)& 5(1)& 4(1)\\\hline
\end{tabular}
\end{center}
\end{table*}

\begin{table*}
\caption{\label{3} The amplitude, $H_y(x_i)$ (scaled by a factor of $10^{1}$), of the flux distribution at each consecutive transverse plane $x_{i}$ from the quarks forming the base, $A$, of isoscoles triangle. The measurements for base source separation distance $A=0.6, 0.8$ and $1.0$ fm for the temperature $T/T_c = 0.9$ are indicated as a function of the third quark position, $Q_{3}$.}
\begin{center}
\begin{tabular}{ccccccccccccccc}\hline
Plane& $x=1$&$x=2$&$x=3$ & $x=4$ &$x=5$&$x=6$&$x=7$&$x=8$& $x=9$ & $x=10$ & $x=12$  & $x=13$  \\ 
  $Q_{3}=R/a$  \\ \hline\hline

A=0.6 fm\\
07&2.73(1)&2.83(1)&2.77(1)&2.58(1)&2.27(2)&1.88(2)\\
08&2.71(1)&2.79(0)&2.75(1)&2.62(1)&2.42(2)&2.14(2)&1.79(2)\\
09&2.68(0)&2.73(0)&2.67(0)&2.56(0)&2.41(1)&2.24(1)&2.01(2)&1.71(1)\\
10&2.64(0)&2.64(1)&2.55(1)&2.43(0)&2.30(0)&2.19(1)&2.06(1)&1.89(1)&1.63(1)\\
11&2.59(0)&2.55(2)&2.42(2)&2.27(1)&2.14(0)&2.05(0)&1.99(1)&1.91(1)&1.77(1)&1.56(1)\\
12&2.53(1)&2.45(3)&2.29(3)&2.10(2)&1.95(2)&1.86(1)&1.83(1)&1.81(0)&1.77(1)&1.67(1)&1.49(1)\\
13&2.46(2)&2.35(3)&2.15(3)&1.93(3)&1.76(3)&1.66(3)&1.63(2)&1.64(1)&1.64(0)&1.63(1)&1.56(2)&1.43(1)\\
A=0.8 fm\\
07&2.88(2)&2.93(1)&2.85(0)&2.64(0)&2.33(1)&1.94(1)\\
08&2.84(2)&2.86(2)&2.79(0)&2.63(0)&2.41(1)&2.14(1)&1.81(1)\\
09&2.80(3)&2.77(3)&2.68(1)&2.52(0)&2.35(0)&2.18(1)&1.97(1)&1.70(1)\\
10&2.74(2)&2.66(4)&2.53(3)&2.36(1)&2.20(0)&2.07(0)&1.96(1)&1.82(1)&1.60(1)\\
11&2.67(2)&2.54(5)&2.38(4)&2.18(3)&2.00(2)&1.89(1)&1.83(0)&1.78(1)&1.69(1)&1.51(2)\\
12&2.59(1)&2.44(5)&2.23(4)&2.00(4)&1.81(3)&1.69(2)&1.64(1)&1.64(0)&1.63(1)&1.58(2)&1.46(2)\\
13&2.51(1)&2.35(4)&2.11(4)&1.85(4)&1.64(4)&1.50(3)&1.45(2)&1.46(1)&1.50(1)&1.52(2)&1.50(3)&1.42(3)\\
A=1.0 fm\\
07&2.85(5)&2.84(4)&2.78(2)&2.59(0)&2.31(0)&1.96(0)\\
08&2.79(6)&2.74(5)&2.68(3)&2.52(1)&2.33(0)&2.10(0)&1.82(1)\\
09&2.72(6)&2.61(6)&2.52(4)&2.36(2)&2.20(1)&2.06(0)&1.91(0)&1.69(1)\\
10&2.64(6)&2.46(7)&2.33(5)&2.15(3)&1.99(2)&1.89(0)&1.83(0)&1.74(1)&1.58(1)\\
11&2.55(6)&2.31(7)&2.14(6)&1.93(5)&1.75(3)&1.66(1)&1.63(0)&1.63(0)&1.60(1)&1.48(1)\\
12&2.46(5)&2.19(7)&1.98(6)&1.73(5)&1.54(4)&1.43(2)&1.41(1)&1.44(0)&1.49(1)&1.49(1)&1.43(2)\\
13&2.39(5)&2.12(7)&1.86(6)&1.59(5)&1.38(4)&1.25(3)&1.21(2)&1.25(1)&1.32(1)&1.39(2)&1.43(3)&1.38(4)\\\hline

\end{tabular}
\end{center}
\end{table*}
\begin{table*}
\caption{\label{4} The squared width, $r_y^{2}$, in lattice units, of the flux distribution as in Table~\ref{2}.}
\begin{center}
\begin{tabular}{cccccccccccccccc}\hline
 plane& $x=1$&$x=2$&$x=3$ & $x=4$ &$x=5$&$x=6$&$x=7$&$x=8$& $x=9$ & $x=10$ & $x=12$  & $x=13$  \\ 
  $Q_{3}=R/a$&  \\ \hline\hline
A=0.6 fm\\

07&13.2(1)&12.8(1)&12.5(0)&12.0(1)&11.4(2)&10.9(3)\\
08&13.5(1)&13.0(1)&12.9(0)&12.6(1)&12.1(1)&11.4(2)&10.7(3)\\
09&13.8(1)&13.2(1)&13.2(0)&13.0(0)&12.7(1)&12.1(2)&11.4(2)&10.6(3)\\
10&14.2(1)&13.3(2)&13.4(1)&13.4(0)&13.2(0)&12.8(1)&12.2(2)&11.3(2)&10.5(3)\\
11&14.6(1)&13.3(2)&13.5(2)&13.6(1)&13.6(1)&13.4(1)&13.0(1)&12.2(2)&11.2(2)&10.2(3)\\
12&15.1(2)&13.3(3)&13.5(2)&13.7(2)&13.9(2)&13.9(1)&13.7(1)&13.1(1)&12.1(2)&10.9(2)& 9.8(2)\\
13&15.6(2)&13.2(3)&13.4(3)&13.7(3)&14.0(3)&14.3(3)&14.3(2)&13.9(1)&13.1(1)&11.9(2)&10.5(2)& 9.3(2)\\

A=0.8 fm\\

07&16.6(4)&15.8(3)&15.1(1)&14.3(0)&13.3(2)&12.4(3)\\
08&16.9(4)&16.1(3)&15.6(1)&14.9(0)&14.0(1)&13.0(2)&12.0(3)\\
09&17.3(4)&16.3(4)&15.9(2)&15.4(1)&14.7(0)&13.8(1)&12.7(2)&11.7(3)\\
10&17.9(5)&16.5(5)&16.2(3)&15.8(2)&15.4(1)&14.7(1)&13.7(2)&12.5(2)&11.3(3)\\
11&18.5(5)&16.7(5)&16.4(4)&16.2(3)&15.9(2)&15.5(1)&14.7(1)&13.5(2)&12.2(3)&10.9(3)\\
12&19.3(5)&16.8(6)&16.7(5)&16.6(4)&16.5(3)&16.2(2)&15.7(2)&14.7(1)&13.4(2)&11.8(3)&10.5(3)\\
13&20.0(5)&17.0(6)&17.0(5)&17.0(5)&17.0(4)&16.9(4)&16.6(3)&15.9(2)&14.7(2)&13.1(3)&11.4(3)& 9.9(3)\\

A=1.0 fm\\
07&21.4(9)&20.1(7)&18.8(3)&17.4(0)&15.9(1)&14.5(3)\\
08&21.9(9)&20.4(8)&19.3(5)&18.1(2)&16.7(0)&15.2(2)&13.8(3)\\
09&22(1)  &20.7(9)&19.7(6)&18.7(3)&17.5(1)&16.1(1)&14.5(2)&13.0(3)\\
10&23(1)  &21(1)&20.2(7)&19.3(5)&18.3(2)&17.1(1)&15.6(1)&13.9(2)&12.4(3)\\
11&24(1)  &22(1)&20.7(9)&19.9(6)&19.1(4)&18.1(2)&16.8(1)&15.1(2)&13.4(3)&11.7(3)\\
12&25(1)  &22(1)&21.4(1)&20.8(7)&20.1(5)&19.2(4)&18.1(3)&16.5(2)&14.7(2)&12.8(3)&11.2(4)\\
13&27(1)  &23(1)&22.4(1)&22.0(8)&21.4(7)&20.7(5)&19.7(4)&18.2(3)&16.4(3)&14.3(3)&12.3(4)&10.5(4)\\\hline

\end{tabular}
\end{center}
\end{table*}

\subsection{Flux Radius profile}
\begin{figure*}[!hpt]
\includegraphics[width=16.5cm]{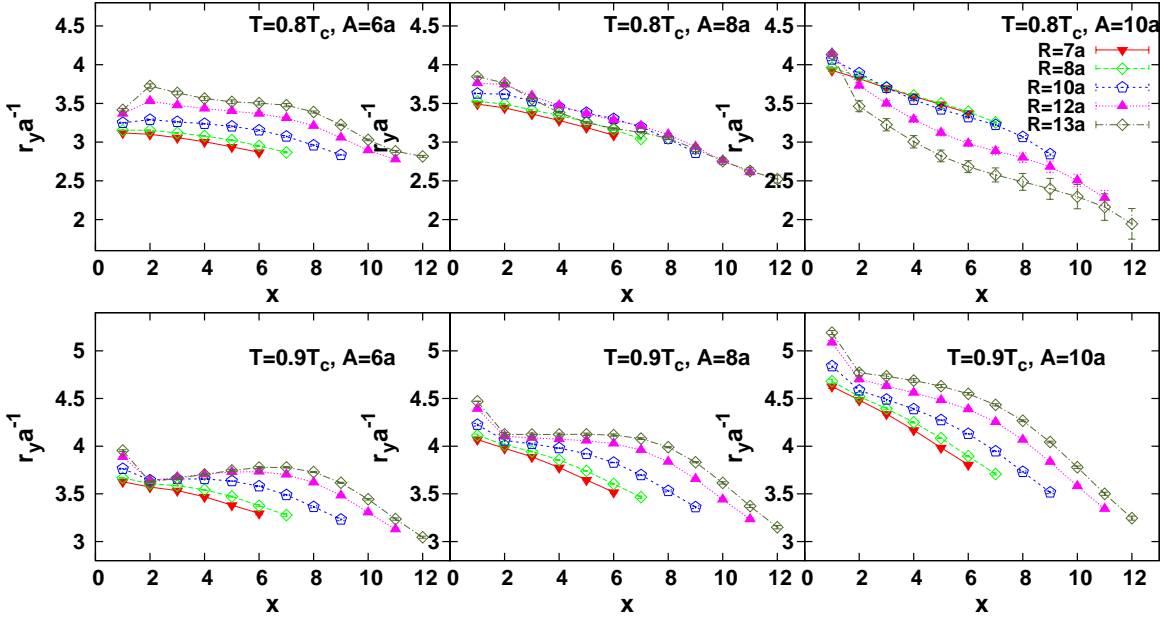}
\caption{\label{profileradius1} The radius profile of the flux-tube measured in the plane of the quarks for each isosceles configuration with base $A=6\,a$, $A=8\,a$ and $A=10\,a$ ($a=0.1$ fm), at two temperature $T/T_c=0.8 $ (top) and $T/T_c=0.9$ (bottom). The legend (in the upper right corner graph) signifies the third quark position.}
\end{figure*}
 At large quark separation, the revealed flux tube profile using the 3Q Wilson loop operator at zero temperature exhibits a uniform tube amplitude with a radius that is only slightly increasing up to the position of the junction~\cite{Bissey}. Although the bias of the revealed energy distribution by the shape of the configurations of the spatial links~\cite{Bissey,Okiharu2004745} leaves these rendered energy distributions somewhat uncertain, this flux distribution has been considered consistent with the parametrization of the 3Q ground state potential with a $Y$-ansatz at large distance~\cite{Takahashi:2000te}. The Y-shaped gluonic distribution  has also been considered in consonance with the dual superconductivity picture~\cite{PhysRevD.11.970,thooft,Mandelstam1976245} of the QCD vacuum. The flux is squeezed into a thin region dual to the Abrikosov vortex~\cite{Cardaci} resulting in the formation of Y-shaped string-like flux tube~\cite{SST95,SSTI95,CH95,QC01}. 

  At finite temperature, on the other hand, one intuitively would expect the quantum vibrations of the underlying three string system~\cite{Jahn2004700, PhysRevD.79.025022, PhysRevD.70.074506} to give rise to a nonuniform action density distribution in a similar fashion to the results revealed in the meson~\cite{PhysRevD.82.094503}. The thin string-like Y-shaped flux tube may delocalize away from its classical configuration and span the whole region though out the bulk of the triangular 3Q configuration, giving rise to a rounded concentric family of $\Delta$ action iso-surfaces (equi-action surfaces of Fig.~\ref{isosurf}). Each surface is weighted by a temperature-dependent  amplitude intensity distribution. In this non-uniform action density context, the radius topology is not fixed merely based on the distribution of Equi-action surfaces, as there can be an infinite number of iso-surface topologies of the action density that all correspond to the same measured square root of the second moment of the distribution. 

  The second moment, $r_y^{2}(x)$, and the amplitude, $H_y(x)$, of the flux density at each line $\vec{\rho}(x_i,y,0)$ is measured by means of Gaussian fits to the complementary distribution $\mathcal{C'}=1-C$ 

\begin{equation}
\mathcal{C'}(\vec{\rho}(x_i,y,0))=H_y(x_i)\,e^{-y^{2}/2r^{2}}
\label{fits}
\end{equation}

  The fits to this Gaussian form are illustrated in Fig.~\ref{gaussb}. The mean square width in the 3Q plane at position, $x_i$, is measured via
\begin{equation} 
\label{widthg}
 r_y^{2}(x_i)=  \dfrac{\int \, d\,y\,  y^{2} \,\mathcal{C'}(\vec{\rho}(x_{i},y,0))} {\int \, d\,y \,\mathcal{C'}(\vec{\rho}(x_{i}, y, 0)) },
\end{equation}
 eliminating dependence on , $H_y(x_i)$. The values of the measurements of , $H_y(x_i)$ and $r_y(x_i)$ are listed in Tables~\ref{1} through ~\ref{4}. The radius profile in the quark plane, $z=0$, is measured at each lattice co-ordinate $x_{i}$. The data points corresponding to radii along the $x$-axis for a given quark configuration are interpolated with a continuous line up to the third quark, $Q_3$, with position $\vec{\rho}(R,0,0)$ as in Fig.~\ref{profileradius1}.

   The first row of graphs in Fig.~\ref{profileradius1} correspond to radii measurements at the temperature $T/T_c=0.8$ with base length running from $A=0.6$~fm to $A=1.0$~fm. For $A=0.6$~fm, the radius profile draws almost constant lines with small declination indicating a subtle decrease along the x-axis up to the third quark position. The difference in radii between the very first planes and the planes close to the third quark $Q_3$ becomes more pronounced with the increase of the third quark $Q_3$ separation $R$ as well as the increase of the distance between the two quarks $Q_{1,2}$ in the base. 

   At the same temperature scale $T/T_c=0.8$ and small isosceles base $A=0.6$~fm, the tube's radius,  $r_y(x_i)$, at a given point broadens slowly with the increase of the quark separation $Q_{3}$. This behavior changes as the length of the isosceles base $A$ becomes wider. The change in radius along the $x$-axis with $R$ approaches near a stagnation in the broadening for $A=0.8$~fm indicating an inflection point. 
\begin{figure}[!hpt]
\includegraphics[width=6.5cm]{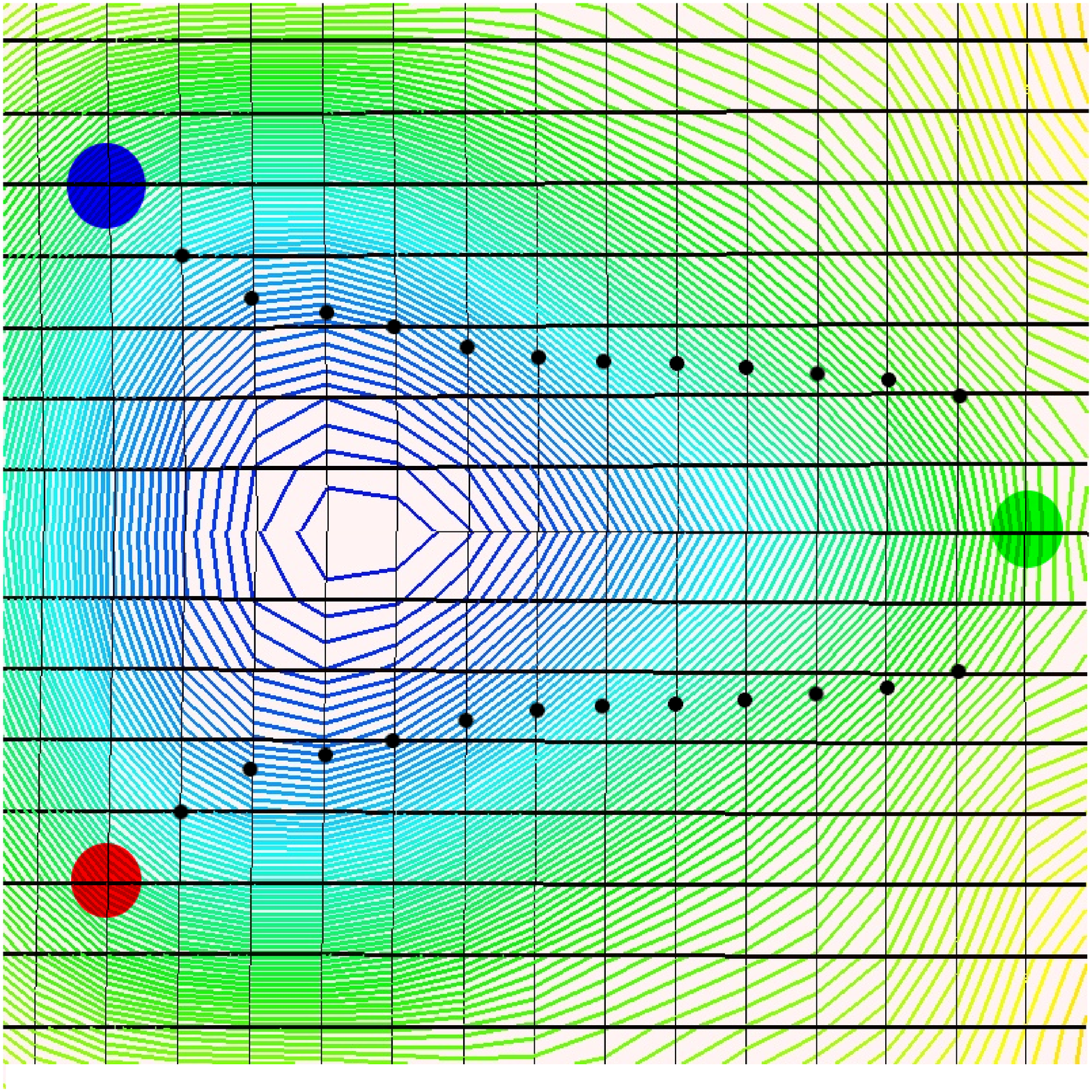}
\caption{\label{radcon} The radius profile of the flux-tube displaying a Y-shape like profile for quark configurations of base $A=1.0$~fm and the third quark position $R=1.3$~fm at temperature $T/T_c=0.8$. In the background is the corresponding flux action-density contours.}
\end{figure}

\begin{figure}[!hpt]
\begin{center}
\includegraphics[width=8cm]{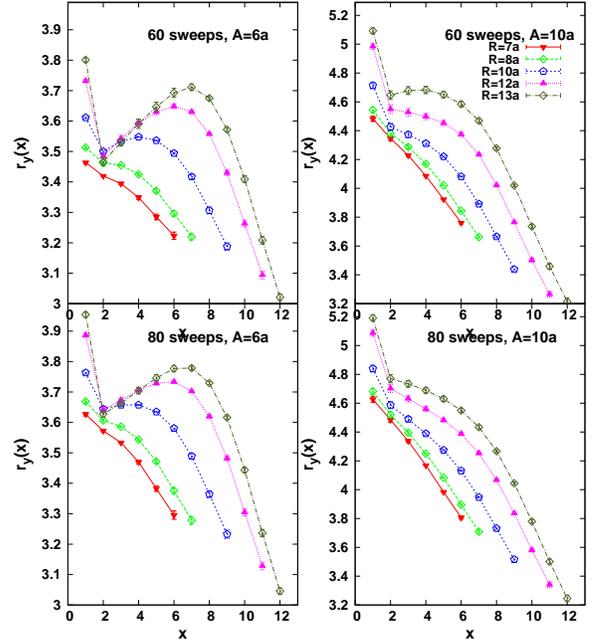}
\caption{\label{smearradius} Same as Fig.~\ref{profileradius1} for isosceles configuration bases of $A=0.6$~fm, and $A=1.0$~fm. The top and bottom figures compare the measured radius profile for two levels of smearing, 60 sweeps and 80 sweeps, respectively. The radius in lattice units.}
\end{center}
\end{figure}

\begin{figure*}
\begin{center}
\includegraphics[width=14.5cm]{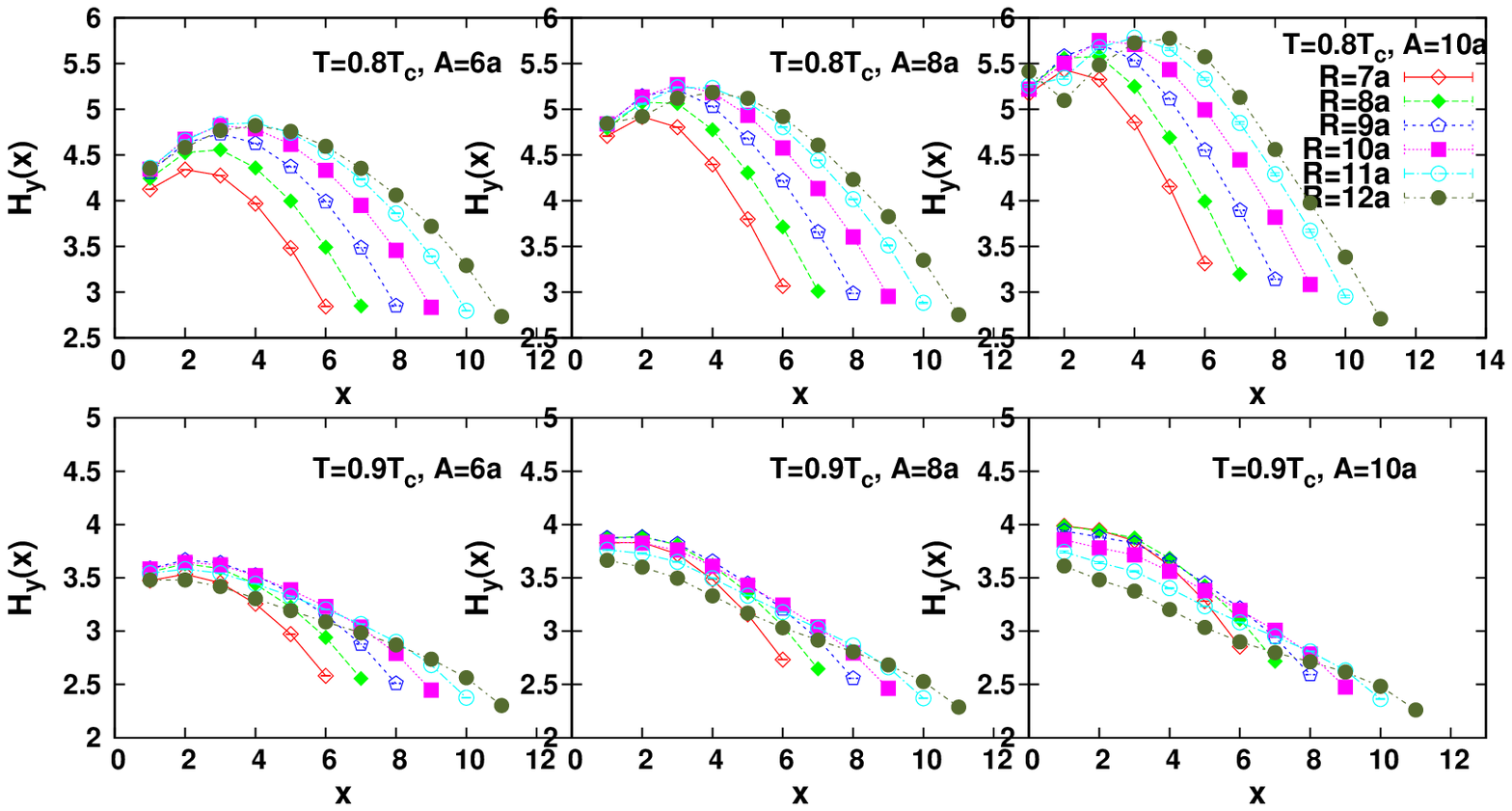}
\caption{\label{profileamp} The profile of the action density amplitude,$H_y(x_i)$ (scaled by a factor of $10^{1}$) for each isosceles configuration with base $A=0.6$ fm, $A=0.8$ fm and $A=1.0$ fm, for the two temperatures $T/T_c=0.8\,$ (top), $T/T_c=0.9\,$ (bottom). The legend signifies the third quark position.}
\end{center}
\end{figure*}

 This is evident from the profile at the widest base length $A=1.0$~fm. The radius at a given point typically decays with the increase of the third quark $Q_3$ separation $R$. The geometrical area spanned by the triangle made up by the 3Q system becomes significantly large as the third quark $Q_3$ steps farther away. In response, the gluonic energy condenses in narrower extents around the $x$-axis. In fact, we see a clearly identifiable Y-shaped profile of the gluon flux emerging at $R=1.3$ fm as well as at $R=1.2$ fm. 

  For illustration, the action contours and radius profile have been superimposed in Fig.~\ref{radcon}. The action contours are concentric convex $\Delta$-shaped action isolines, the corresponding radii measured along the $x$-axis are not coinciding with any of the action isolines and have a convex Y-shaped-like profile. Variation of the amplitude, $H_y$, hides the underlying Y-shape revealed through the consideration of $r_y$. We see from the corresponding last graph of the bottom row in Fig.~\ref{profileradius1} that the contour lines and the radius profile have similar concave curvatures at the temperature $T/T_c=0.9$. 

This analysis may render the widely used terms such as the $Y$ and $\Delta$ shaped gluon flux linguistically ambiguous if their usual usage is brought to the regime of non-uniform action density profiles with position-dependent (local) amplitudes distribution, $H_y$. One can speculate that the observation of the simultaneous co-existence of both the $Y$ and the $\Delta$ aspects of the profile opens the possibility that the ground-state baryon state may exhibit a similar action isosurface behavior even with the success of the Y-ansatz in the parametrization of the large distance potential. This may sound plausible especially if we take into account the fact that the gluonic junction broadens with the evolution of Euclidean time in the Wilson loop operator~\cite{Bissey}. In addition, non-uniformly UV regulated Wilson loop operators have optimized the ground state overlap at zero-temperature in the mesonic sector~\cite{Bakry:2011cn}. At the same time, and along the line of the above argument, the observation of a Y-shape current distribution following abelian gauge fixing~\cite{Bornyakov} at $T/T_c=0.8$ should not be taken by no means as contradictory to the $\Delta$-shaped action density in QCD observed without gauge fixing.       

  Inspection of the bottom row of graphs in Fig.~\ref{profileradius1} reveals how thermal effects on the tube's radius profile take place as we get closer to the deconfinement point at $T/T_c=0.9$. In general, there is an increase in the radius of the flux with the increase of the temperature. The tube's topology is almost the same with an expansion of the size as we get to wider triangular bases. The radii flatten out through the planes $x \leq 6$ for large quark separations. The change in radius along the x-axis increases also with the increase of the temperature for small quark separations. Minimum growth in the radius for increasing $R$ is noticeably manifesting near the Fermat point of the configurations $x=1.7,\, 2.3$ and $2.9$ for the isosceles bases $A=0.6, 0.8$ and 1.0 fm, respectively. Another distinguishable feature for the profile at $T/T_c=0.9$ is that the radius does not show any sign of squeezing at all quark configurations. The increase in energy resulting from the increase of the temperature is now large enough to accommodate the corresponding enlargement in the geometrical area of the triangle set up by the quarks. We focus on detailed aspects of the flux broadening patterns separately in Sec.~D.              

  In addition to the force measurements in Sec.~\ref{Forces} taken as a guiding analysis to set a trusted distance scale for each level of smearing, we now report the effects of smearing on the radius profile of the action density along the tube. Fig.~\ref{smearradius} compares the radii of the flux at each plane $x$ measured on 60 and 80 sweep smeared gauge configurations. The values of the measured radii do not change at distant planes from the isosceles base. Smearing causes a subtle shifting rather than lensing effect on the radius at the planes near the quarks in the base $Q_{1,2}$. An increment of 20 sweeps of smearing from 60 sweeps to 80 sweeps causes a maximum increase of the radius by a subtle factor of $1.04$. This effect diminishes as we consider far planes $x\,>\,6$ from the $Q_{1,2}$ quarks on the base.

\subsection{Flux Amplitude profile} 

  At zero temperature, the revealed vacuum structure inside the static baryon constructed via Wilson loop operator has a maximum vacuum suppression at the center of the triangle made up by the 3Q system at small separations~\cite{Bissey}. At large distances, the Wilson loop operator of the minimum spatial string length has been found to minimize the potential~\cite{Bissey}, indicating a junction position at the Fermat point of the configuration. However, a peak in the action density at zero temperature does not manifest. The distribution assumes a constant amplitude. The analysis performed here for the density distributions using Polyakov lines as hadronic operators, nevertheless, reveals density amplitude peaks which manifest at short as well as large source separation distances. In this section, the amplitude profile is investigated and contrasted for the two considered temperatures.


  Tables~\ref{1} and~\ref{3} summarize the measured amplitudes, $H_y(x)$, in accord with the Gaussian fits of Eq.~\eqref{fits}. The corresponding plots are shown in Fig.~\ref{profileamp}, for the isosceles configurations with base $A=0.6\,$ fm, $A=0.8$ fm and $A=1.0$ fm at two temperatures $T/T_c=0.8$ (top row) and $T/T_c=0.9$ (bottom row), respectively. At all considered planes, the height of the distribution $H_y(x)$ decreases with the increase of the temperature, which reciprocates the changes of the radius of the flux with the temperature. The decrease of the distribution height together with the associated increase in the distribution moment indicates the spread of the gluonic energy with the increase of the temperature.

   At $T/T_{c}=0.8$, the amplitude also increases at most planes when moving the third quark $Q_3$ farther from the base of the isosceles configuration. Recalling the corresponding decrease in the radii along the x-axis, one infers the gluonic behavior undergoes a localization rather than a decay of the flux tube. The amplitudes at  $T/T_{c}=0.9$ show similar increase up to small quark separation. However, a noticeable turn over to decreasing amplitude with the increase of the third quark $Q_3$ separation manifests for $R \geq 10\,a$. The behavior of the amplitude and radius at $T/T_{c}=0.8$ resembles respectively the behavior of the radius and amplitude at $T/T_{c}=0.9$. The analysis of the flux amplitude shows different qualitative behavior as we transit from the end of the QCD plateau to just before the deconfinement point and this behavior is reciprocal to the radius profile indicating a delocalization of the gluonic distribution with the increase of the temperature and the subsequent decrease in the string tension. 

   In addition to the position of amplitude maximum along the $x$-axis at $y=0$ and the corresponding trigonometric aspects of the triangular setup. The maxima localize around the second and third planes $x=2,x=3$, for third quark $Q_{3}$ separations $R<10\,a$. However, this localization of the maxima of the vacuum suppression around the Fermat points ceases as the third quark is pulled away further. The density maximum moves in the same direction of third quark at $T/T_c=0.8$ and moves in the opposite direction (towards the triangle base) for the higher temperature $T/T_{c}=0.9$.    

\subsection{\label{growth} The broadening of the flux width} 
\begin{figure}[!hpb]
\begin{center}
\includegraphics[width=7.5cm]{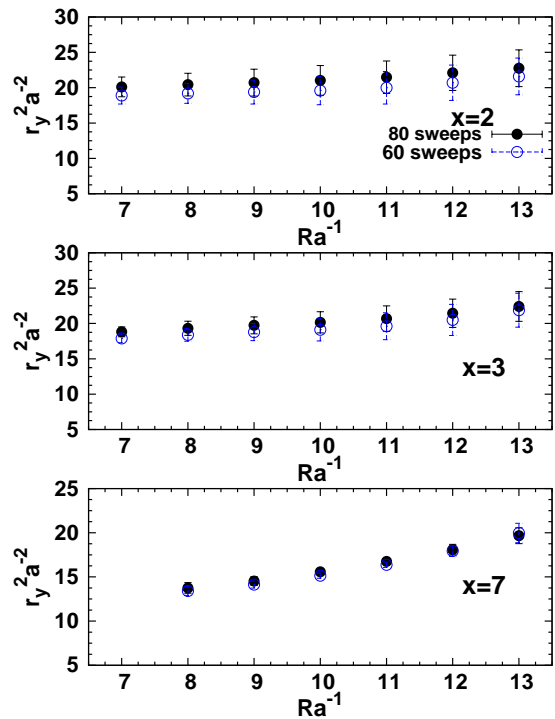}
\caption{\label{compsmear} The squared flux distribution width at the depicted planes, $x_i=2,3$ and 7, are compared for two smearing levels. The isosceles configuration base length $A=1.0$ fm at temperature $T/T_c=0.9$. Smearing merely shifts the profile by a constant. The broadening pattern is not affected.}
\end{center}
\end{figure}
\begin{figure*}[!hpt]
\begin{center}
\includegraphics[width=18cm]{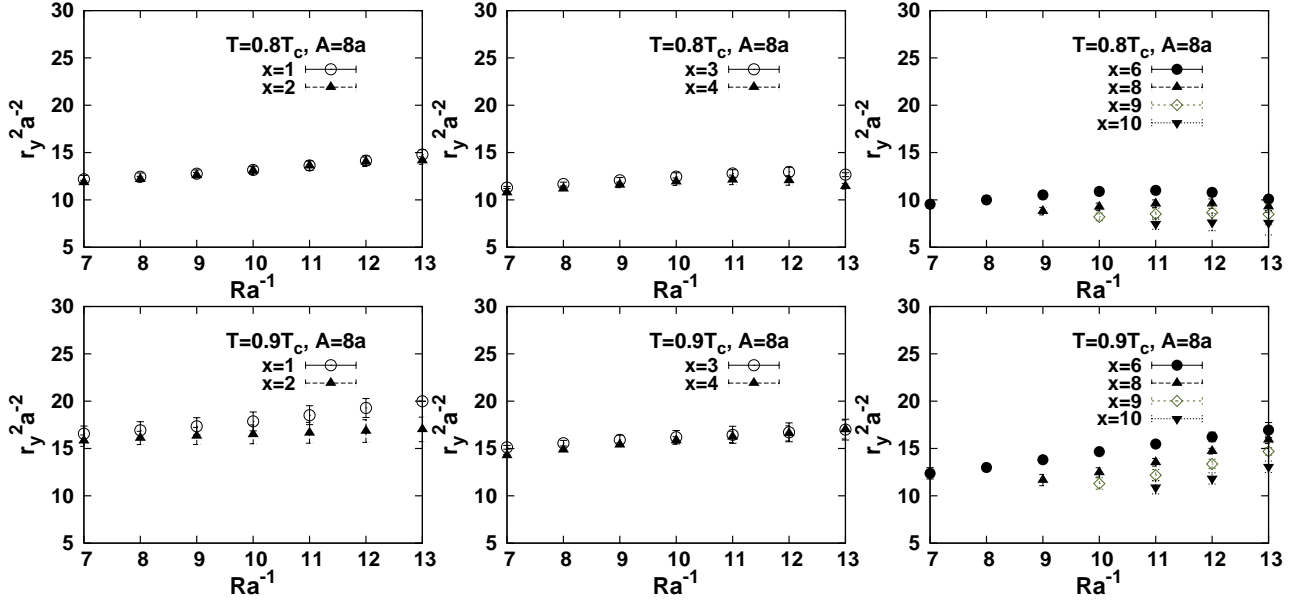}
\caption{\label{compa8t89} The squared flux-tube width at the depicted planes for the isosceles configuration $A=0.8$ fm compared at two temperatures $T/T_{c}=0.8$ (top) and $T/T_{c}=0.9$ (bottom). The plane coordinates are indicated in the legend.}
\end{center}
\end{figure*}   

\begin{figure*}[!hpt]
\begin{center}
\includegraphics[width=18cm]{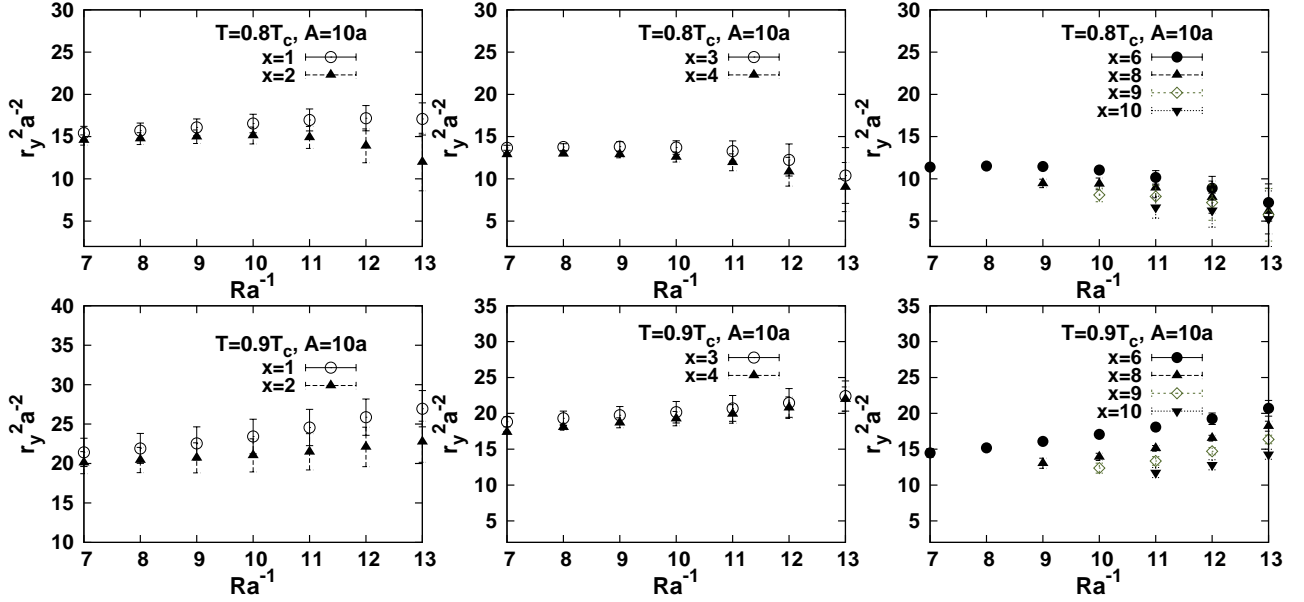}
\caption{\label{compa10t89} Same as Fig.~\ref{compa8t89} for a larger isosceles base length of $A=1.0$ fm.}
\end{center}
\end{figure*}   

\begin{figure*}[!hpt]
\begin{center}
\includegraphics[width=19cm]{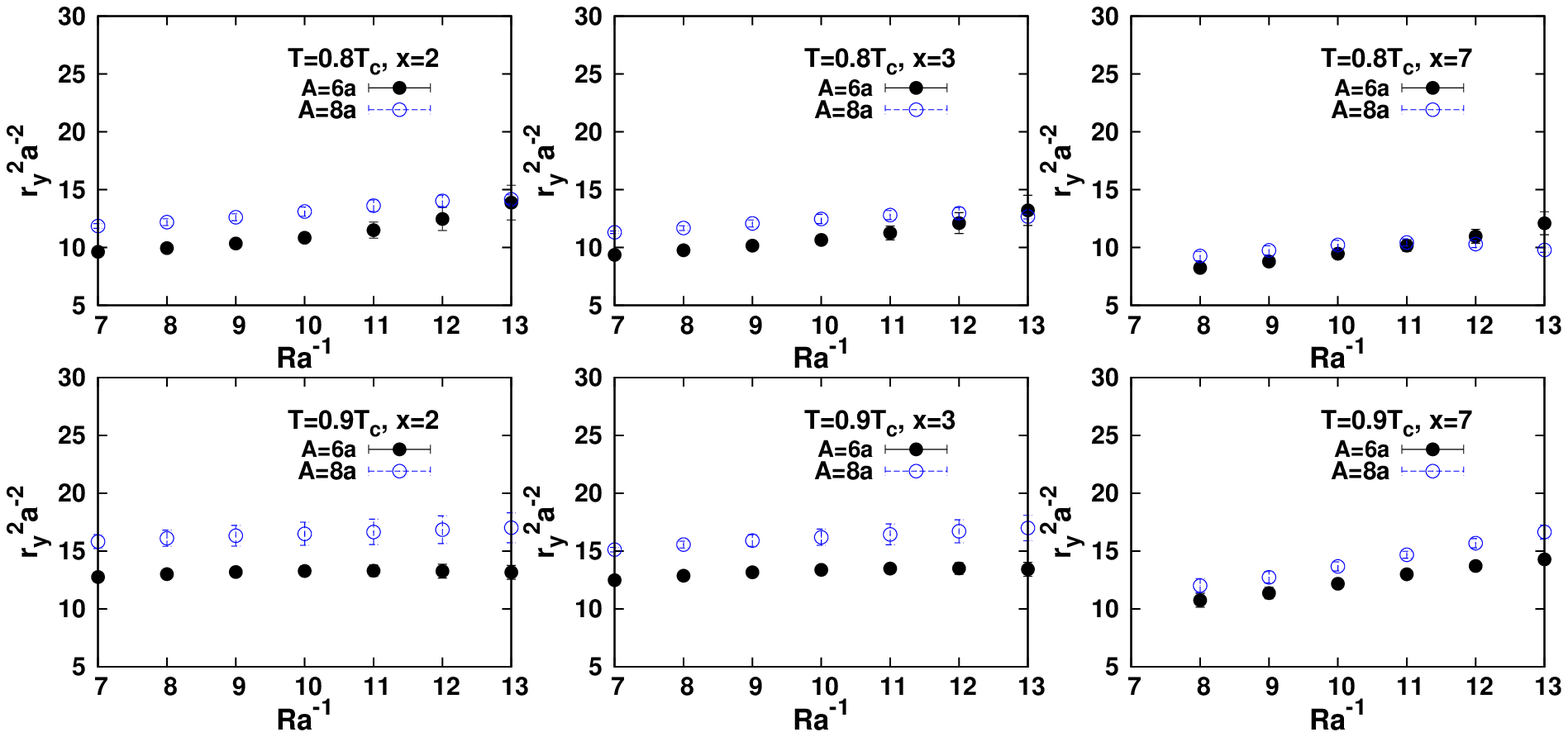}
\caption{\label{68} Comparison of the mean square width of the flux distribution at three distinct planes, $x= 2, 3,$ and 7, for two isosceles bases $A=0.6\,$ fm and $A=0.8$ fm. The upper graphs show the comparison at $T/T_c=0.9$ whereas the lower are at $T/T_c=0.8$.}
\end{center}
\end{figure*}

    In this section, we focus on the broadening aspects of the mean-square width of the flux. We restrict our analysis to the mean square width in the 3Q plane at the two considered temperature scales. The lattice data for the mean square width, $r_y^2(x_i)$, at planes $x_i$ along the $x$-axis are summarized in Tables~\ref{1} and~\ref{3}. For convenience, we have considered the tube's width for an analysis performed on gauge configurations of 80 sweeps of smearing, where we obtain the best signal to noise ratio with only a relatively small elimination of short distance points which are affected by smearing. 

  In the last section, we reported the effects of gauge-smoothing on the radius of the gluon flux. Smearing shifts the width by a subtle constant near the base of the isosceles triangle, as in Fig.~\ref{compsmear}. This shift diminishes at distant planes from the base. To further examine the rate of broadening of the flux distribution, we fit the mean square width to the simple linear ansatz  

\begin{equation} 
\label{linear}
r_y^2(R;x_i)=b_1(x_i)\,R + b_2(x_i)
\end{equation}
\begin{table}[!hpt]
\caption{\label{chi} The slope of the growth in the mean square width, $r_y^2$, measured for isosceles base $A=1.0$ fm  
on two levels of link smearing. The measurements are obtained from the fits to the linear form Eq.~\eqref{linear}.}
\begin{ruledtabular}
\begin{tabular}{cccc}
 $n_{\rm{sw}}$ & $b_1$& Fit Range $Ra^{-1}$ \\
\hline
x=2  \\
   60  &0.76(4)& 4-13  \\
   80  &0.83(3)& 4-13   \\
x=7  \\
   60  &2.3(1) & 8-13   \\
   60  &2.8(2) & 10-13  \\ 
   80  &2.2(1) & 8-13   \\
   80  &2.5(1) & 10-13
\end{tabular}
\end{ruledtabular}
\end{table}

\noindent The returned values of the slope of the growth in the flux width with the increase of the isosceles height, R, display small systematic errors associated with the selection of parameters as indicated in Table~\ref{chi}. 
  
   The profile of the broadening of the glue at various planes, for isosceles base, $A=0.8$ fm, is plotted in Fig.~\ref{compa8t89} with a similar plot for, $A=1.0$ fm, in Fig.~\ref{compa10t89}, respectively. Each set of data describes how the width of the gluonic flux vary at a given plane $x_i$ for the triangle base as the third quark $Q_3$ moves to a larger values of $R$. Evidently, the increase of the temperature dramatically increases the rate of the broadening of the glue at all planes. 

   Apart from the pronounced thermal effects near the deconfinement point, we see the rates of broadening at $T/T_c=0.8$ are decreasing as one proceeds to the more distant planes from the base of the triangle. Moreover, the wider the base of the isosceles triangle, the more pronounced is the corresponding decrease in the width, indicating that the gluonic tend to become more localized as the geometrical area enclosed by the quarks positions becomes larger. 

 The shrinking of the width of the flux tube is a peculiar property to certain geometrical configurations of the Multi-quark system. The decrease in the width with the increase of the inter-quark separation has never been observed in the meson either using Polyakov lines at finite temperature~\cite{PhysRevD.82.094503} or Wilson's loop at zero temperature~\cite{Bali:1994}. The analysis of the Wilson loop based energy distribution at zero temperatures does not seem to indicate shrinking of the width of the flux tube~\cite{Bissey}.  

\begin{figure*}[!hptb]
\begin{center}
\includegraphics[width=16cm]{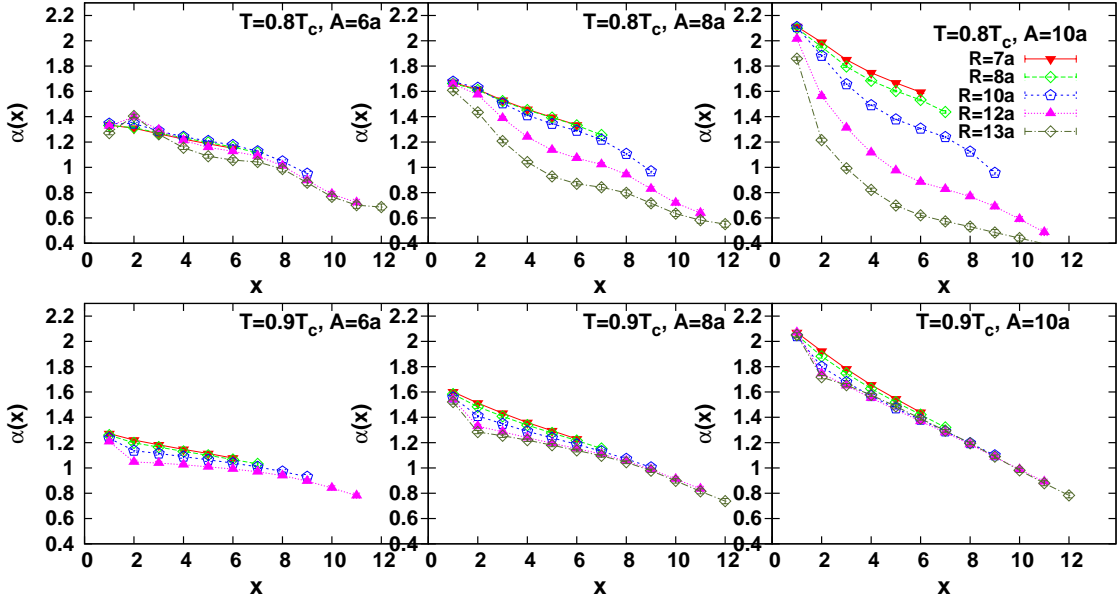}
\caption{\label{asp} Comparison of the ratio, $\alpha$ (Eq.~\eqref{alpha}), of the mean squared width of the flux distribution parallel and perpendicular to the quark plane for three isosceles bases $A=0.6\,$ fm, $A=0.8$ fm and $A=1.0$ fm. The upper graphs show comparison at $T/T_c=0.8$ whereas the lower at are $T/T_c=0.9$.}
\end{center}
\end{figure*}
 
  Near the deconfinement point, the broadening of the mean square width, $r_y^2$, exhibits a clear linear divergence at distant planes from the base of the isosceles triangle for large separations, $R>1.0$ fm of the third quark $Q_3$.  This result resembles the observed linear growing in the flux distribution width at the same temperature in the meson~\cite{Bakry:2010sp}. The slope of the growing width at distant planes from the base, $x>6$, from plane to plane show only subtle changes. This indicates that the effects of the boundary and the junction fluctuations decays away by proceeding to large quark separations. Fig.~\ref{68} displays the effects of increasing the distance between the two quarks $Q_{1,2}$  on the rate of change of the width versus the motion of the source $Q_3$. At $T/T_c=0.9$ the wider the base the faster the rate of growth. This behavior is the reciprocal of the corresponding one at $T/T_c=0.8$.   


\subsection{\label{growth}Planes aspect ratio}
   The gluonic flux in the 3Q system does not exhibit a symmetry between the width measured in the quark plane and that in the perpendicular direction. This has to do with the underlying gluonic structure and the associated fluctuations. For example, the string picture indicates an asymmetry in the mean square width between the two planes~\cite{PhysRevD.79.025022}. We report for completeness the general qualitative features of this ratio of the action density in the two perpendicular planes. The width of the tube in the perpendicular direction is measured through Gaussian fits as

\begin{equation}
\label{widthgz}
  r_z^{2}(x_i)=  \dfrac{\int \, d\,z\,  z^{2} \,\mathcal{C'}(\vec{\rho}(x_{i},0,z))} {\int \, d\,z \,\mathcal{C'}(\vec{\rho}(x_{i}, 0, z)) }.
\end{equation}

\noindent We measure the ratio between the width in the quark plane and that in the perpendicular plane to the quarks 

\begin{equation}
\label{alpha}
  \alpha(x_i) \equiv \frac{r_x^2(x_i)}{r_z^2(x_i)}. 
\end{equation}

\noindent It is interesting to consider this quantity since the predictions of the string model at zero temperature indicate asymmetric width pattern near the junction~\cite{PhysRevD.79.025022}. In a similar way to the characteristics of the flux that we have studied in the previous sections, we plot in Fig.~\ref{asp} the aspect ratio at the two temperatures for the same quark position configurations. 

  Generally, the value of the aspect ratio indicates that the fluctuations in the quark plane are always larger than the perpendicular fluctuations for both temperatures. Further inspection of Fig.~\ref{asp} shows only subtle dependence on the temperature for the small isosceles bases. The aspect ratio is changing slowly as we move through the planes up to the third quark position for the smallest isosceles bases. However at larger bases, the asymmetry throughout the gluonic cone becomes pronounced. The results of the aspect ratio are indicating greater restoring forces for the gluonic distribution in the perpendicular direction to the quark plane. This effect diminishes as we consider planes away from the Fermat point of the triangle quark configuration. Near the deconfinement point $T/T_c=0.9$, the aspect ratio increases with the base length showing only a subtle dependency on $R$. This behavior is reversed at the lower temperature $T/T_c=0.8$ for large quark separation. Indeed we see at $T/T_c=0.8$, for the bases $A=0.8$ fm and $A=1.0$ fm, a decrease in the aspect ratio at all planes with the increase of the third quark separation $R>1.0$ fm. 

  Within the underlying gluonic picture of three squeezed string-like flux tubes meeting at a junction and assuming a Y-shaped form, the predictions of this string model for the ratio of the mean square width of the flux distribution in the quark planes and the perpendicular direction at position of the junction has been worked out in Ref~\cite{PhysRevD.79.025022}. However, one should carefully consider the geometrical aspects of the configuration and take into account any remnant ($T/T_c=0.8$) or pronounced ($T/T_{c}=0.9$) thermal effects before pursuing a complete confrontation with the predictions of the baryonic bosonic string models. We included the thermal effects~\cite{bakry:178} into the predictions that have been worked out earlier in Ref.~\cite{PhysRevD.79.025022}. However, this comparison of the lattice data with the predictions of the models lies beyond the scope of the present presentation of the gluonic profile, and will be reported in details elsewhere~\cite{Bakr}.

  Finally, it is worth noting that the flux strength distribution revealed with the action density using the Wilson loop does not appear to produce an asymmetric gluonic pattern. For instance, in Ref.~\cite{Bissey} the radius of the tube is calculated with cylindrical coordinates assuming a cylindrical symmetry of the tube. The analysis provided here for the aspect ratio of the mean-square width provides another distinct feature of the glue, as revealed by Polyakov loops rather than a manifestation of temperature effects.          
                                  
\section{\label{conc}Conclusion}
   The gluon flux distribution of a three quark system in pure SU(3) Yang-Mills vacuum has been revealed at finite temperature. This analysis is an extension of the calculations of the action density correlations obtained recently for the $Q \bar{Q}$~\cite{PhysRevD.82.094503} system to three quark systems. The infinitely heavy baryonic state has been constructed by three Polyakov loops. The gluon flux is measured as a correlation between the action density operator and three traced (gauge-invariant) Polyakov lines. Measurements have been taken near the end of QCD phase diagram, $T/T_c \approx 0.8$ and just before the deconfinement point $T/T_c \approx 0.9$.\\ 

{\bf Noise reduction}~has been achieved using a gauge-independent statistical approach that exploits the space-time symmetries as well as symmetries of the quark configuration space. The calculations are performed at each point of the lattice and averaged over the lattice four-volume. The average over the configuration space have been calculated on 500 independent bins. Each bin consists of an average over measurements that are closely spaced in configuration space. An over-improved version of the stout-link smoothing algorithm has been employed with number of sweeps such that the physics is preserved in a systematic and controlled manner. 

  We have revealed the characteristics of the flux action-density measured for three sets of geometrical 3-quark configurations and the corresponding changes on the behavior due to the temperature. Each set corresponds to isosceles triangle bases of length $A=0.6$ fm, $A=0.8$ fm and $A=1.0$ fm. The characteristics of the isosurface, the radius and the amplitude profiles of the action density correlations, in addition to the broadening (or the shrinking) pattern of the flux distribution, can be summarized in the following main points:  

{\bf A.~The Iso-surface}~of the flux action-density displays a family of concave $\Delta$ shapes at small as well as large quark separations. These $\Delta$-shaped gluonic distributions persist and do not change into a Y-shape as the distances between the quark sources are increased. The density plots in the quark plane display a nonuniform distribution at all distance separations. This contrasts the Wilson loop results at zero temperature which exhibit uniform action density along each arm of the Y-shaped profile. A remarkable feature of the revealed map of the contour lines of the flux strength is that the shape of the contour lines do not show significant sensitivity to the temperature for the two temperatures considered here.\\

{\bf B.~The Radius profiles}~give indications on the spread of the energy inside the baryon. At the lowest temperature near the end of the plateau, $T/T_c=0.8$, the measurements of the radius indicate localization of the action density in narrow regions for quarks separations greater than 1.0 fm. The radius of the tube decreases and draws a Y-shaped like profile even though the action isosurface and isolines are $\Delta$ shaped. Near the deconfinement point, on the other hand, the energy tends to spread as we see the radius increases at all considered distance scales.    

{\bf C.~The Amplitude profile}~analysis of the flux density shows a maximum vacuum fluctuations suppression at the plane nearest to  the Fermat point of  the planar three-quark configurations for intermediate separation distances. The distribution's peak ceases localizing around the Fermat point of the 3Q isosceles configurations when the height, $R$, is greater than $1.0$ fm. The peak shifts to the outside of the triangle made at $T/T_c \approx 0.9$ and shifts in the reverse direction to the inside of the triangle for $T/T_c \approx 0.8$. That is the amplitude gets higher when the radius shrinks at $T/T_c \approx 0.8$ and the reverse manifest at $T/T_c \approx 0.9$.        

{\bf D.~The Flux mean-square width}~does not always broaden with the increase of the quark source separation as is the case in the meson. For the lowest temperature, $T/T_c \approx 0.8$, the flux distribution shrinks in width for large quark separations. The change in the width of the flux tube shows a non-broadening aspect which is a property of certain configurations of the multi-quark system. The width, however, grows linearly near the deconfinement point, $T/T_c \approx 0.9$,  with the increase of the height of the triangle. In general, the slope of the decrease or increase in the width, at both temperatures, depends on the length of the triangle base. The wider the base of the triangle set up by the quarks positions, the lower or higher is the slop at temperatures $T/T_c=0.8$ and $T/T_c=0.9$, respectively.  


{\bf E.~The Aspect ratio} between the mean square width of the flux distribution in the quark plane and the width in the perpendicular plan exhibits an asymmetry. The gluonic fluctuations in the plane of the quarks are greater than that in the perpendicular directions around Fermat point, indicating a greater restoring force for the system in the plane of the quarks. The ratio between the two components of the mean square width decreases as we consider planes farther than the locus of the Fermat point of the quark configuration. The temperature dependence for the aspect ratio is more pronounced at large quark separations while we see almost the same profiles for small isosceles bases. The deviation of the aspect ratio from unity is implied by the predictions of the string models and do not manifest using Wilson loop operator in the action correlations.            

{\bf Future work} \\
  As this work presents our first investigation of the flux distribution of the 3Q system at finite temperature, there are many promising avenues of investigation remaining. For example, additional quark configurations could be examined in detail. Lower temperatures remain to be investigated. We compare ~\cite{Bakr} the growth of the mean square width of the flux distribution measured on the lattice with finite-temperature extensions of the baryonic string model ~\cite{bakry:178,Bakr} for the width of the baryonic junction ~\cite{PhysRevD.79.025022}. On can also study the relevant ansatz for the measured potentials at each temperature. Methodological improvements based on increasing the number of measurements and decreasing the number of gauge smoothing sweeps are always desirable. The method pursued here may prove effective in a calculation framework that includes the effects of the dynamical quarks. 
\section*{Acknowledgments}
 This research was undertaken on the NCI National Facility in Canberra, Australia, which is supported by the Australian Commonwealth Government. We also thank eResearch SA for generous grants of supercomputing time which have enabled this project. This research is supported by the Australian Research Council.


\appendix

\end{document}